# Vibration-Based Damage Detection in Wind Turbine Blades using Phase-Based Motion Estimation and Motion Magnification

Aral Sarrafi, Zhu Mao*, Christopher Niezrecki, Peyman Poozesh
Department of Mechanical Engineering
University of Massachusetts Lowell
One University Avenue, Lowell, MA 01854
zhu_mao@uml.edu, 978-934-5937

## ABSTRACT

Vibration-based Structural Health Monitoring (SHM) techniques are among the most common approaches for structural damage identification. The presence of damage in structures may be identified by monitoring the changes in dynamic behavior subject to external loading, and is typically performed by using experimental modal analysis (EMA) or operational modal analysis (OMA). These tools for SHM normally require a limited number of physically attached transducers (e.g. accelerometers) in order to record the response of the structure for further analysis. Signal conditioners, wires, wireless receivers and a data acquisition system (DAQ) are also typical components of traditional sensing systems used in vibration-based SHM. However, instrumentation of lightweight structures with contact sensors such as accelerometers may induce mass-loading effects, and for large-scale structures, the instrumentation is labor intensive and time consuming. Achieving high spatial measurement resolution for a large-scale structure is not always feasible while working with traditional contact sensors, and there is also the potential for a lack of reliability associated with fixed contact sensors in outliving the life-span of the host structure. Among the state-of-the-art non-contact measurements, digital video cameras are able to rapidly collect high-density spatial information from structures remotely. In this paper, the subtle motions from recorded video (i.e. a sequence of images) are extracted by means of Phase-based Motion Estimation (PME) and the extracted information is used to conduct damage identification on a 2.3-meter long Skystream® wind turbine blade (WTB). The PME and phased-based motion magnification approach estimates the structural motion from the captured sequence of images for both a baseline and damaged test cases on a wind turbine blade. Operational deflection shapes of the test articles are also quantified and compared for the baseline and damaged states. In addition, having proper lighting while working with high-speed cameras can be an issue, therefore image enhancement and contrast manipulation has also been performed to enhance the raw images. Ultimately, the extracted resonant frequencies and operational deflection shapes are used to detect the presence of damage, demonstrating the feasibility of implementing non-contact video measurements to perform realistic structural damage detection.

## 1 Introduction

Structural health monitoring (SHM) has been widely applied in a variety of applications over the past several years with a goal of improving reliability, achieving lower cost condition-based maintenance, performing damage prognosis, and eliminating unexpected catastrophic failures [1, 2]. Several approaches have been adopted for structural damage detection, such as vibration testing [3], ultrasonic guided wave inspection [4, 5], acoustic emission [6, 7], thermal imaging[8], etc. In all SHM efforts, the identification of the healthy state of the structure is the very first step that later yields to a decision-making process to identify the presence of damage after processing the later measurements [9]. After data cleansing and normalization, the measurements undergo the process referred to as the feature extraction that preserves the damage sensitive properties of the acquired data and eliminates the redundant information that are not contributing to the decision-making process [9]. The decision-making procedure is then performed based on the extracted features. This stage oftentimes is accomplished by a human in the loop as an expert in order to detect the occurrence of the damage based on the extracted features. The process may also be handled automatically using machine learning, statistical data processing or other autonomous classification, and clustering algorithms [10].

Vibration-based SHM follows the above-mentioned procedure, i.e. the dynamic response of the structure is measured by means of instruments such as accelerometers or strain sensors, and the damage-sensitive features being used in vibration based SHM are extracted (e.g. resonant frequencies, damping, mode shapes or operating deflection shapes (ODS), or other dynamic sensor signal features) [3]. That being said, experimental and operational modal analysis (EMA and OMA) are the two widely adopted techniques used in structural dynamics identification to extract features employed in vibration-based SHM and the corresponding decision-making procedure [3], as well as being used for model updating and model validation in several applications related to aerospace, civil and mechanical systems [11, 12].

For the vibration-based SHM approach, decision-making regarding the status of the system is made according to the changes in dynamic behavior of structures [13]. Sensor systems play a major role in both EMA and OMA approaches as well as other activities for structural dynamics identification. An ideal sensor system for modal analysis should be able to record the vibration information of the structure accurately with high spatial resolution without alternating the dynamics of the structure induced by mass loading effects or changes to the stiffness. Accelerometers are precise sensors having a broadband frequency range of measurement and are relatively inexpensive, therefore are utilized widely for modal analysis. However, instrumentation of lightweight structures with accelerometers induces mass loading effects, and can make unwanted changes to the dynamics of the structure. Despite the efforts of compensating the mass loading effects, the accuracy of numerous mass cancelation methods is not satisfactory [14, 15]. Although avoiding mass loading effects in system identification procedure (EMA and OMA) is a valuable advantage, in SHM context the mass loading effect is not a major concern because both the damaged and baseline cases are affected by the induced changes due to the accelerometers. In addition, performing modal testing on large-scale structures with accelerometers can be highly time consuming and labor intensive because the installation of a large number of accelerometers on the structure and handling all the wiring and instrumentation is not always straightforward [16]. More importantly, current accelerometers are only able to measure the responses at discrete locations, and are not able to provide full-field spatial resolution, which is necessary in many situations such as mode shape curvature-based SHM techniques [17]. The maintenance of large-scale infrastructures equipped with accelerometers or other conventional discrete sensors, is also costly and this is another major drawback of using accelerometers or strain gages, especially for long-term SHM applications.

Laser vibrometers are capable to record the response of structures without any mass loading effects and do not alter structural stiffness [18, 19]. In addition, high spatial resolution may also be achieved if the laser vibrometers are empowered with precise scanning capabilities. However, laser scanning vibrometers are relatively expensive and the response of the structure is measured sequentially which increases the measurement time considerably if a high spatial distribution is required. Likewise, for structures with very large or low-frequency displacements, the scanning laser measurement systems are not always effective.

Digital video cameras combined with image and video processing algorithms provide another non-contact approach for structural measurements [20-26]. Compared to laser vibrometers, digital cameras are low-cost and capable to make simultaneous full-field measurements. Recently, high resolution and high-speed (frame rate) digital video cameras in conjunction with stereo-photogrammetry approaches, such as 3D digital image correlation (3D DIC) and 3D point tracking (3DPT), have been utilized to perform modal analysis and strain analysis on large-scale industrial structures [27-34]. Although 3D DIC and 3DPT provide smooth mode shapes with high accuracy, these methods generally require a speckle pattern or high contrast markers mounted or painted on the surface of the structure. Surface preparation for large-scale structures could be enormously time-consuming or even not practical if the structure has an extreme operating environment. Considering the high computational burden of 3D DIC and 3DPT, processing the collected image data in real time is not currently viable except for low frequencies. As an alternative, phase-based computer vision algorithms have recently gained more attention for structural motion estimation, and system dynamics identification to avoid the above-mentioned surface perpetration and decrease the size of data associated with 3D DIC and 3DPT that may lead to less computational effort as well. In 2015, Chen *et al.* [35-37] utilized phase-based motion estimation (PME) combined with a motion magnification algorithm to extract the resonant frequencies and operational deflection shapes of several lab scale benchmark structures, including a cantilevered beam and a pipe test specimen. The main advantage of PME is that surface preparation is no longer necessary for the analysis, and the algorithm is capable of extracting the approximate motion information using the natural features of the structure or images given normal lighting conditions. The computation consumption of PME is also likely to be less than traditional 3D DIC and 3DPT, because PME only acquires images from single camera. Moreover, finding the matching points in the right and left cameras is a main source of computational cost in 3D DIC and 3DPT that is not required in the PME technique. It should be also highlighted that 3D DIC and 3D point tracking are mature and reliable techniques and PME still needs further investigations to be able to compete with the above mentioned traditional techniques. Yang *et al.* [38] applied PME and motion magnification in conjunction with blind source separation, generating an unsupervised learning method for modal analysis. The applied blind source separation algorithm is able to extract the modal parameters automatically, so that user supervision and calibrating the input parameters are avoided. The same authors also investigated the effect of sub-Nyquist sampling on video measurements and the PME algorithm [39]. It has been also shown that using phase-based motion magnification as a pre-processing stage can enhance the signal-to-noise ratio in specific frequency bands for 3D DIC and 3DPT [40]. Moreover, preliminary results of modal identification on a wind turbine blade using PME and motion magnification can be found in [41]. Currently, most of the research conducted on the applications of PME and video magnification are limited to lab scale structures yet the capabilities of this technique for damage detection is not evaluated as well. Thus far, PME has not been applied to full-scale industrial structures with complicated geometry and material properties.

In this paper, PME and video magnification are employed to perform operational modal analysis on a 2.3-m long wind turbine blade (WTB) and the resonant frequencies and operating deflection shapes (ODS) are extracted. Monitoring the condition of the wind turbine blades is important to help reduce excessive maintenance costs and to be able to predict premature failure [42] . Structural damage is also introduced to the WTB by attaching an external mass to represent change in mass distribution. The same PME procedure is applied on the damaged WTB to identify the deviations in dynamic behavior of the structure from its undamaged baseline. The approach to extract the modal parameters associated with baseline and damaged WTB provided by PME and motion magnification is valuable to identify mass/stiffness changes such as ice forming on the edge of a WTB or wing. In this work, the boundary conditions provided are identical to the realistic scenario when wind turbine blades are attached to the hub in operations. The proposed methodology is able to identify the dynamics of the non-rotating structures, and it can be considered as a routine inspection technique for wind turbine blades while the turbine is not rotating. For operating wind turbines, de-rotating algorithms need to be adopted to process the acquired videos before applying PME for structural dynamics identification.

The paper starts with a brief literature review and theoretical background of PME that is presented in section 2. Formulation of the PME and simulation results are presented for a simplified special case to establish a foundational intuition of the methodology employed in this paper. The underlying approach of phase-based motion magnification is also discussed briefly in section 2. The experimental setup is then described in section 3 including detailed specifications of the test structure (WTB), image acquisition system, and the testing procedure for both the baseline and the damaged conditions. The results provided by PME and motion magnification are then processed for decision-making regarding the condition of the WTB. The occurrence of damage, such as icing of field turbine blades, is simulated by adding mass to the blade tip, and is detected in the modal domain. In more detail, the resonant frequencies of the baseline and damaged WTB are extracted using PME. Issues associated with poor lighting has been addressed by manipulating the histogram of the sequence of images, in order to extract the operating deflection shapes and achieve a better visualization. Within this paper, non-contact Phase-based Motion Estimation and Magnification are used for the first time to identify the dynamics of a real-world wind turbine blade having complex geometry to demonstrate that practical damage detection can be performed.

# 2 Theoretical Background on Phase-Based Motion Estimation (PME) and Video Magnification

Motion estimation from a sequence of images is an active research area in computer vision. The very first attempts to estimate the motion field in videos can be traced back to the intensity-based optical flow estimation techniques, such as Lucas-Kanade (LK)[43], and Horn-Schunck (HS)[44] methods. Optical flow is defined as the relative motion field between an object and observer, and intensity based optical flow is calculated by solving the aperture equation at each pixel, which contains the spatial and temporal derivatives of the sequence of images. Several applications of the intensity based optical flow method in structural dynamics identification can be found in [45, 46]. The main drawback of these intensity-based optical flow approaches is their sensitivity to noise and disturbances, such as abrupt changes in the illumination of the scene, which induces error to the estimated motion field. Fleet *et al.* [47, 48] introduced the Phase-based Motion Estimation (PME) method in order to overcome some of the difficulties that intensity-based motion estimation techniques are not able to handle while estimating the motion field from a sequence of images. PME is able to extract the motion by analyzing the phase variations instead of manipulating the raw pixel intensity values. It has been shown that phase variations are robust to noise and disturbances and as a result, the PME outperforms other intensity-based motion estimation techniques [38, 49].

## 2.1 Phase-Based Motion Estimation

In this section, the PME algorithm utilized in this paper is reviewed and is used to estimate the motion of a vibrating WTB that has complex geometry. This paper suggests a modified and simplified version of the PME algorithm proposed by Fleet *et al.* [47], and the theoretical approach of the conduction of motion estimation and magnification is reviewed in [50]. The PME mainly originates from the Fourier transform shift theorem, which indicates that the any motion in the spatial domain results in variations to the phase in the frequency domain. The kernel function in the Fourier transform is the sinusoid $e^{j\omega x}$ , which has infinite spatial support, therefore the Fourier transform is only able to capture the global motion in any 2D function, and local motion estimation is not possible. In order to estimate local motion, the kernel function should have finite spatial support, such as the 2D Gabor wavelets used in this work, which transform each frame of the video into the complex domain and represent the local motion in a more precise way. The Gabor wavelet is a sinusoid function modulated by a Gaussian envelope, and have similar characteristics to the image processing procedure in human's vision system[51]. The general form for a 2D Gabor wavelet is:

$$g\left(x,y;\lambda,\theta,\psi,\sigma,\gamma\right)=\exp\left(-\frac{x'^2+\gamma^2 y'^2}{2\sigma^2}\right)\exp\left(i\left(2\pi\frac{x'}{\lambda}+\psi\right)\right). \tag{1}$$

As indicated in Equation (1), the Gabor wavelets are functions of *x* and *y*, where the direction of the Gabor wavelet is represented through the variables $x'$ and $y'$ in Equation (2):

$$x'=x\cos\theta+y\sin\theta \quad \text{and} \quad y'=-x\sin\theta+y\cos\theta . \tag{2}$$

In Equations (1) and (2), *x* and *y* are the independent spatial variables and $\theta$ determines the orientation of the Gabor wavelet. In the general form of the 2D Gabor wavelets in Equation (1), $\lambda$ represents the wavelength of the sinusoids, $\gamma$ is the spatial aspect ratio that determines the ellipticity of the Gabor wavelet, $\psi$ is the phase offset, and $\sigma$ is the standard deviation of the Gaussian function modulating the sinusoid. The 2D Gabor wavelet can also be expressed as real and imaginary parts separately with a simpler form:

$$g\left(x,y;\lambda,\theta,\psi,\sigma,\gamma\right)=G_\theta+iH_\theta\,. \tag{3}$$

The real and imaginary pairs for 2D Gabor wavelets are presented for three different orientations, namely 0, 45 and 90 degrees, as shown in Figure 1.

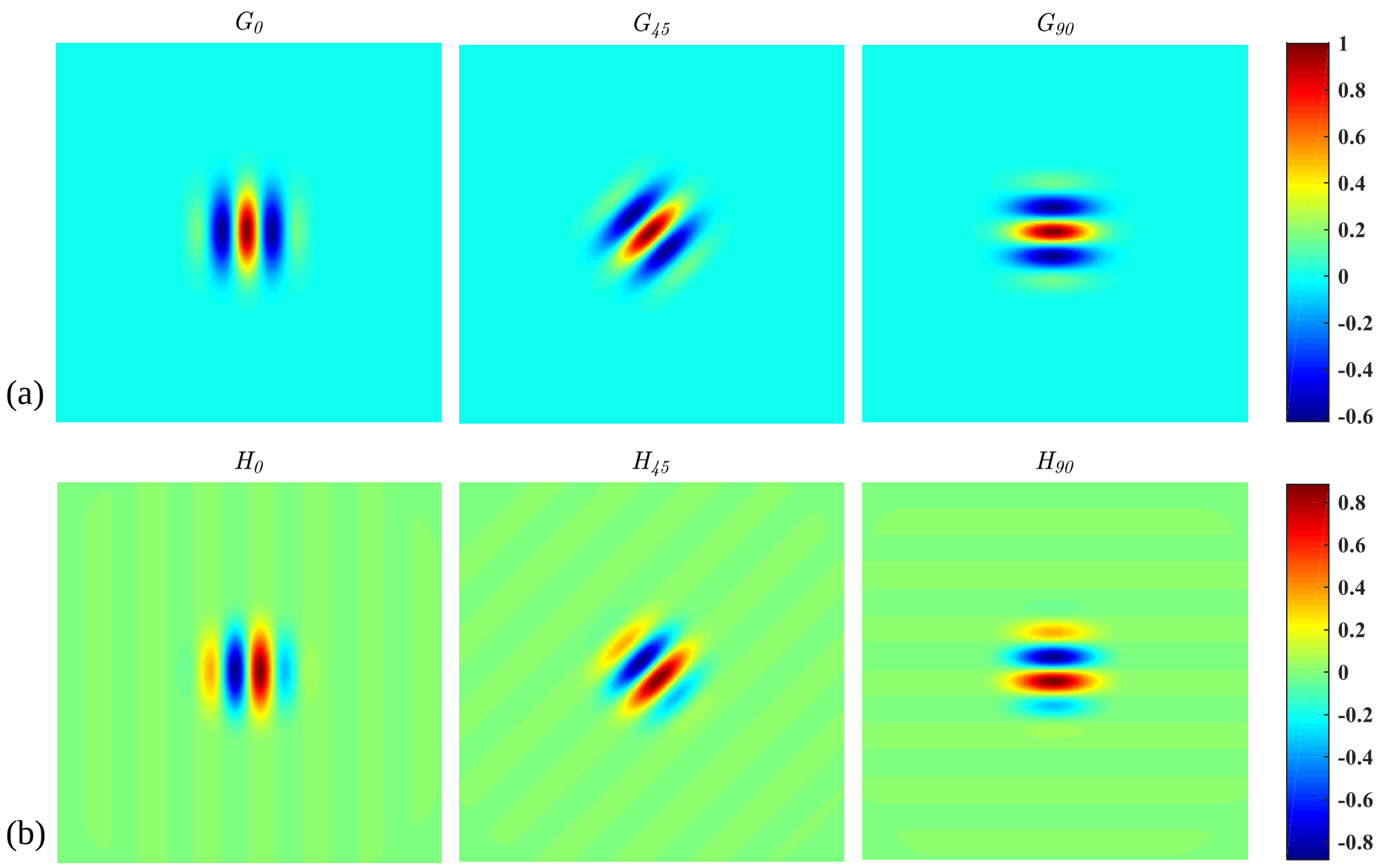

Figure 1. Complex Gabor wavelets in three different orientations; (a) the real parts of complex Gabor wavelets, (b) the imaginary parts of complex Gabor wavelets

A video may be considered as sequence of images $I\left(x,y,t\right)$. Each image at a specific time step is a 2D spatial function $I\left(x,y\right)$ at which *x* and *y* denotes the pixel location in camera sensor plane, and $I\left(x,y\right)$ is the intensity of that pixel (any image is a matrix and each of the elements of that matrix are referred to as one pixel). Assuming that $I\left(x,y,t\right)$ is the image intensity values at time $t$ and location (*x*,*y*), at time $t+\Delta t$, the motion $\left(\delta_x,\delta_y\right)$ is applied to the intensity function, therefore the image intensity for next time step is expressed as $I\left(x+\delta_x,y+\delta_y,t+\Delta t\right)$. It is feasible to assume that the lighting condition does not change during this time window, so that the time dependency of the intensity function could be neglected for simplification. Therefore, the pixel intensities at time $t+\Delta t$ reduces to $I\left(x+\delta_x,y+\delta_y,t\right)$. The formulation of the PME used in this paper is represented for a special case where $\theta=0$ is selected for the Gabor wavelet, and this will reach the estimated motion in *x* direction. Since $\theta=0$, the independent spatial variables can be expressed as $x'=x$ and $y'=y$ according to Equation (2). The approach used in this paper utilizes a transformation $\boldsymbol{T}$ similar to a convolution integral in conjunction with the Gabor wavelets, which decomposes image intensities $I\left(x,y,t\right)$ to a set of complex coefficients at each frame. The transformation $\boldsymbol{T}$ that is used in this paper is defined as in Equation (4):

$$\boldsymbol{T}\left\{I\left(x,y,t\right),g\left(x,y;\lambda,\theta,\psi,\sigma,\gamma\right)\right\}=\int_{-\infty}^{+\infty}\int_{-\infty}^{+\infty}I\left(x,y,t\right)g\left(x-u,y-v;\lambda,\theta,\psi,\sigma,\gamma\right)dxdy\,. \tag{4}$$

In the transformation described in Equation (4) the integration is on *x* and *y*, therefore *x* and *y* can be considered as dummy variables, which will disappear after the integration and the output of the transformation is only a function of *u* and *v*. Therefore, transformation $\boldsymbol{T}$ will map the image intensities at each frame $I\left(x,y,t\right)$ to a new domain $C\left(u,v,t\right)$, with respect to *u* and *v*, as described in Equation (5):

$$C(u,v,t) = \boldsymbol{T}\{I(x,y,t), g(x,y;\lambda,\theta,\psi,\sigma,\gamma)\}. \tag{5}$$

As an example, the Gabor wavelet with $\theta = 0$ is convolved with image intensities at two consecutive frames considering the transformation in Equation (4), and the complex coefficients $C(u,v,t)$ and $C(u,v,t+\Delta t)$ (*x* and *y* are dummy spatial variables, which are used to perform the transformation) are obtained as:

$$C(u,v,t) = \int_{-\infty}^{+\infty}\int_{-\infty}^{+\infty} I(x,y,t)\exp\left(-\frac{(x-u)^2+\gamma^2(y-v)^2}{2\sigma^2}\right)\exp\left(i\left(2\pi\frac{x-u}{\lambda}+\psi\right)\right)dxdy \text{, and} \tag{6}$$

$$C(u,v,t+\Delta t) = \int_{-\infty}^{+\infty}\int_{-\infty}^{+\infty} I(x+\delta_x, y+\delta_y, t)\exp\left(-\frac{(x-u)^2+\gamma^2(y-v)^2}{2\sigma^2}\right)\exp\left(i\left(2\pi\frac{x-u}{\lambda}+\psi\right)\right)dxdy\ . \tag{7}$$

Similar to other space-frequency domain transformations (the independent variable is space), this transformation maps the image intensities at two consecutive frames into the complex domain with coefficients $C(u,v,t)$ and $C(u,v,t+\Delta t)$. Changing the variables of integration to $x+\delta_x = \alpha$ and $y+\delta_y = \beta$ for image intensities at time $t+\Delta t$ in Equation (7), $C(u,v,t+\Delta t)$ yields to:

$$C(u,v,t+\Delta t) = \int_{-\infty}^{+\infty}\int_{-\infty}^{+\infty} I(\alpha,\beta,t)\exp\left(-\frac{(\alpha-\delta_x-u)^2+\gamma^2(\beta-\delta_y-v)^2}{2\sigma^2}\right)\exp\left(i\left(2\pi\frac{\alpha-\delta_x-u}{\lambda}+\psi\right)\right)d\alpha d\beta \tag{8}$$

Equations (6) and (8) may be reorganized in the following format by pulling out the variables in the phase of the Gabor wavelet, which are independent of the integration variables. The new representation is useful to extract the phase:

$$C(u,v,t) = \exp\left(-i\frac{2\pi u}{\lambda}\right)\int_{-\infty}^{+\infty}\int_{-\infty}^{+\infty} I(x,y,t)\exp\left(-\frac{(x-u)^2+\gamma^2(y-v)^2}{2\sigma^2}\right)\exp\left(i\left(2\pi\frac{x}{\lambda}+\psi\right)\right)dxdy \text{, and} \tag{9}$$

$$C(u,v,t+\Delta t) = \exp\left(-i\frac{2\pi}{\lambda}(u+\delta_x)\right)\int_{-\infty}^{+\infty}\int_{-\infty}^{+\infty} I(\alpha,\beta,t)\exp\left(-\frac{(\alpha-\delta_x-u)^2+\gamma^2(\beta-\delta_y-v)^2}{2\sigma^2}\right)\exp\left(i\left(2\pi\frac{\alpha}{\lambda}+\psi\right)\right)d\alpha d\beta\ . \tag{10}$$

The argument of the phase for the terms being integrated are identical for both Equations (9) and (10) (i.e. $\left(2\pi\frac{x}{\lambda}+\psi\right)$ and $\left(2\pi\frac{\alpha}{\lambda}+\psi\right)$), as *x* and $\alpha$ are dummy variables. Therefore, the phase of the definite integrals will be the same, which is denoted as a constant $\phi'$. By comparing Equations (9) and (10), the phase values of complex coefficients in both frames $C(u,v,t)$ and $C(u,v,t+\Delta t)$ may be extracted as below ($\phi'$ is the phase value of the complex number resulting from computing the definite integrals):

$$\phi(u,v,t) = \arg\left[C(u,v,t)\right] = -\frac{2\pi}{\lambda}u + \phi' \text{, and} \tag{11}$$

$$\phi(u,v,t+\Delta t) = \arg\left[C(u,v,t+\Delta t)\right] = -\frac{2\pi}{\lambda}(u+\delta_x) + \phi' \tag{12}$$

It can be shown that the phase difference in two consecutive frames is proportional to the motion, and the motion information is preserved in phase variations (see Equations (11) and (12)):

$$\Delta\phi(u,v) = \phi(u,v,t) - \phi(u,v,t+\Delta t) = \frac{2\pi}{\lambda}\delta_x \text{, or} \tag{13}$$

$$\Delta\phi(u,v) = h\delta_x\ , \tag{14}$$

Where in Equation (14), *h* is a constant coefficient and $h = \frac{2\pi}{\lambda}$. In general, the phase variations are functions of spatial variables (*u*, *v*) or (*x*, *y*), because the motion may be local, and only observed at a limited number of pixels in the video. Therefore, the phase variation has a general formulation with respect to spatial variables, despite the fact that it is constant in this special simulated case study.

Equation (14) implies that motion in $x$ direction, $\delta_x$, may be estimated from the phase variations. Similarly, the motion in $y$ direction will be estimated via changing the orientation of the Gabor wavelet to $\theta = \dfrac{\pi}{2}$ . Selecting other orientations for the Gabor Wavelet will also enables the algorithm to estimate the absolute projected motion in that direction.

Before applying PME to the sequence of images captured from the vibrating wind turbine blade, the performance of the method is evaluated for an artificially contrived simulation case. A 2D-Gaussian surface moving in x-direction is adopted as an artificial sequence of images (video),

$$I\left(x,y,t\right) = A\exp\left(-\frac{\left(x-\delta_x\left(t\right)\right)^2 + y^2}{2s^2}\right),$$

($s$ is the standard deviation of the Gaussian surface and $A$ is an arbitrary constant value) and the applied motion $\delta_x\left(t\right) = e^{-\xi\omega_n t}\sin\left(\omega_n t\right)$ is assigned to the surface to mimic viscously damped oscillations of the Gaussian surface along $x$ axis. As a truncated case, the Gaussian surface in Figure 2 has bounded boundaries therefore the infinite integrals in Equations (6) and (7) can be computed. Figures 2(a) and 2(b) shows the above mentioned Gaussian surface at two instances of time at the peak motion ranges while oscillating with the defined motion $\delta_x\left(t\right) = e^{-\xi\omega_n t}\sin\left(\omega_n t\right)$. The phase difference planes shown in Figure 2(c) moves proportionally to the applied motion as Equation (14) indicates, and the motion may be accurately estimated from the phase variations as shown in Figure 2(d) the estimated motion from phase-based motion estimation (red dots) is very well aligned with the actual motion (blue curve) because the simulated sequence of images are noise free. The same algorithm will be applied on the sequence of images from the vibrating wind turbine blade in order to extract the subtle vibrations of wind turbine blade due to the external impact load for further analysis.

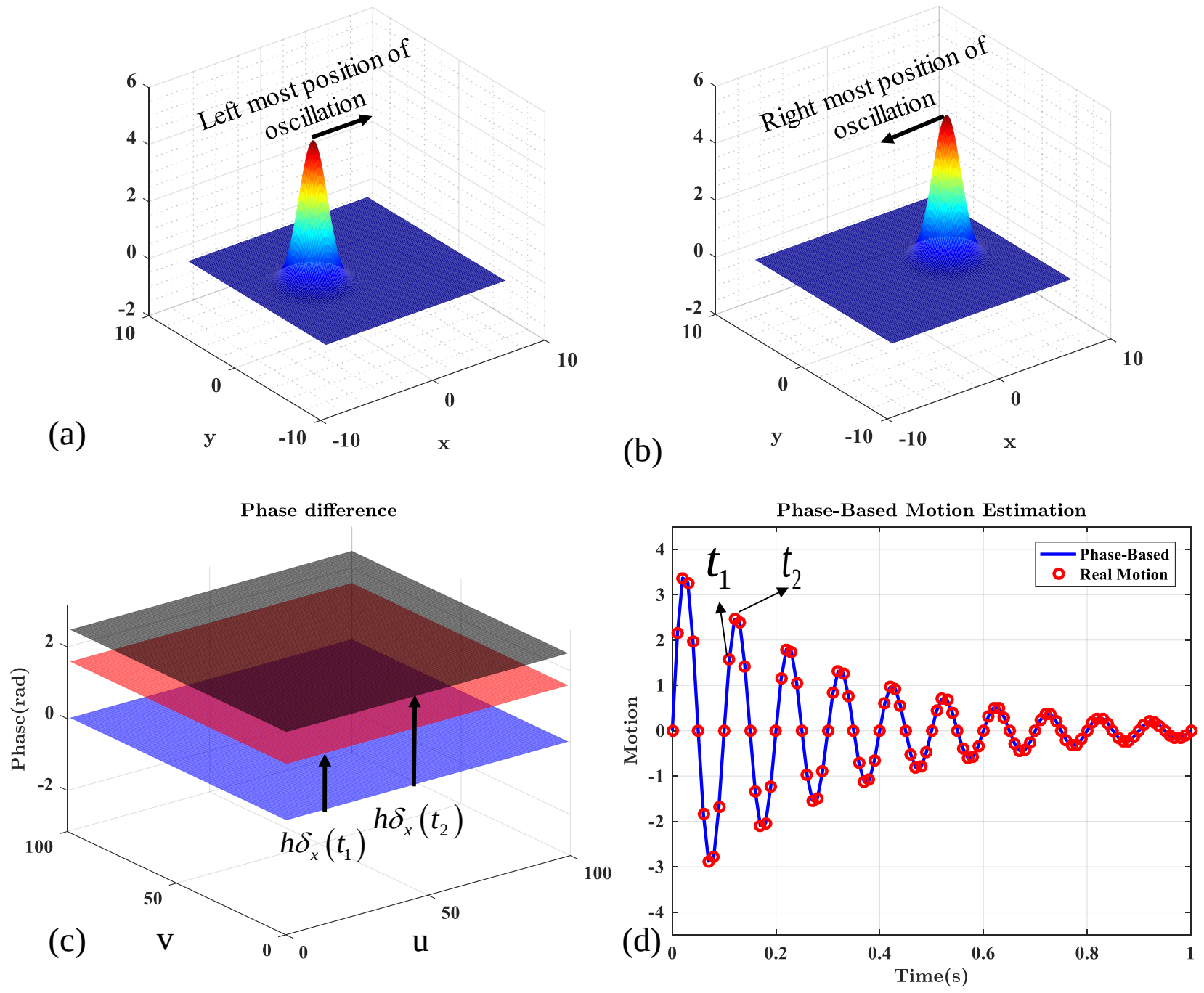

Figure 2. (a) The moving Gaussian surface at its left most position of oscillation; (b) The moving Gaussian surface at its right most position of oscillation; (c) Phase plane varitions preserving the motion information of the moving Gaussian surface; (d) Actual and estimated motion of the Gaussian surface using PME

## 2.2 Phase-Based Motion Magnification

Phase-based motion magnification is an amplified visualization of the motion in specific frequency bands by means of magnifying the phase variations and band-pass filtering the video in the temporal domain [50]. The objective is to manipulate the extracted phase, which contains the motion information, and then reconstruct the video, which magnifies motion present in the video in a specific frequency band. The intensities from the sequence of images $I(x,y,t)$ are transformed to the complex domain $C(u,v,t)$ using transformation in (4) and a 2D Gabor wavelet, as the PME flow described in the last section. The complex coefficients may be expressed as amplitude and phase as $C(u,v,t)=A(u,v,t)e^{i\phi(u,v,t)}$ where $A(u,v,t)$ and $\phi(u,v,t)$ are the amplitude and phase of the complex value at each pixel defined as

$$A(u,v,t)=\|C(u,v,t)\|_2 \text{, and} \tag{15}$$

$$\phi(u,v,t)=\arg\left[C(u,v,t)\right], \tag{16}$$

where ($\|\bullet\|_2$ is the absolute value of the complex number). The time series of phase $\phi(u,v,t)$ may be manipulated by applying an ideal band pass filter **F**(.) with specifications described in Figure 3, in which $f_c$ is the center frequency, $b$ is the width of the pass band, and $\alpha$ is referred to as the magnification factor which determines the amount of motion amplification in the defined frequency band. Manipulating the phase $\phi(u,v,t)$ via filter **F**(.) will result to a new set of phase planes $\Phi(u,v,t)$, which contains the magnified motion within the selected frequency band. In this manipulation, **F**(.) is a digital temporal band-pass filter which is designed to amplify the temporal variations of the phase within a specific frequency band. The general form of digital filters can be expressed as $\mathbf{F}(.)=\dfrac{A(L)}{B(L)}(.)$, where $A(L)$ and $B(L)$ are polynomials ($m>n$) defined as:

$$A(L)=a_0+a_1L+\cdots a_nL^n, \tag{17}$$

$$B(L)=b_0+b_1L+\cdots+b_mL^m, \tag{18}$$

and $L$ is the time lag operator ($L^k\phi(u,v,t)=\phi(u,v,t-k)$). The new time series of phase after filtering can be achieved as:

$$\Phi(u,v,t)=\mathbf{F}\left[\phi(u,v,t)\right]. \tag{19}$$

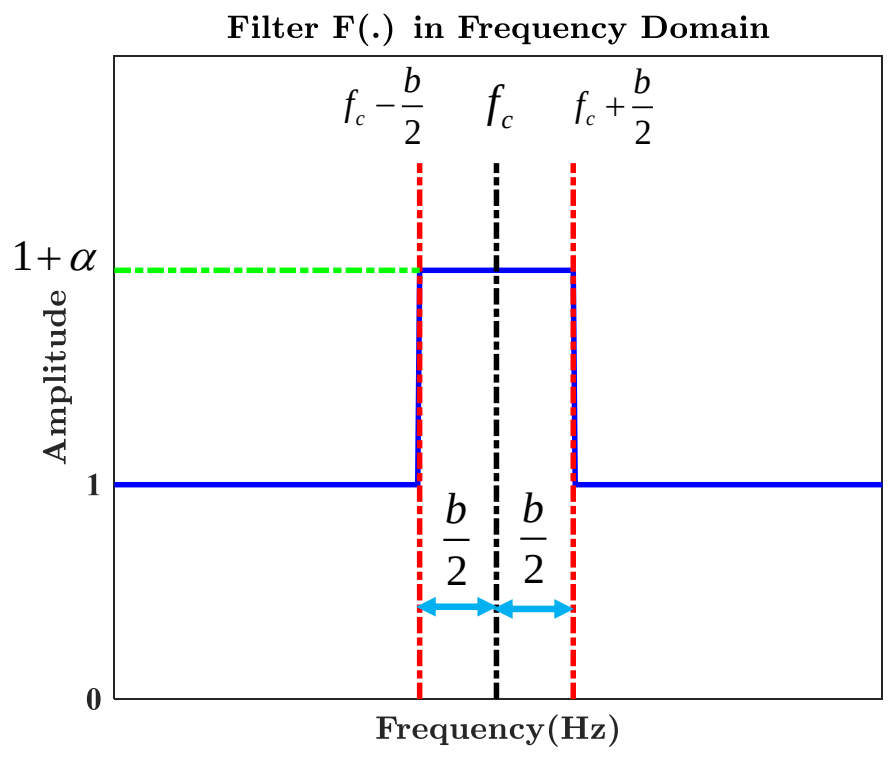

Figure 3. Band pass filter utilized for phase variation manipulation. $\alpha$ , $f_c$ and $b$ are the input parameters for phase-based motion magnification

Applying the band pass filter in temporal domain on phase variations preserves and amplifies the motion information in the specific frequency band, and reconstructing the sequence of images using the filtered phase variations $\Phi(u,v,t)$ will result in a motion-magnified video that visualizes the imperceptible motions in the selected frequency band. Discussion on reconstruction procedure is beyond the scope of this paper and further information regarding this stage can be found in references [49, 50], which is the origin of PME technique. It is worth mentioning that the reconstruction procedure is the inverse transformation that is proposed in Equation (4).

In analyzing the vibrations of structures such as WTBs, if the center frequency is selected to be the resonant frequency of the structure (assuming well separated modes), and $b$ is selected to be small enough to make the frequency band to contain only the resonant frequency of interest, the reconstructed video will represent the operating deflection shape of the structure corresponding to the selected resonant frequency. If the operating deflection shapes (ODS) are reasonably separated in the frequency domain, the magnified sequence of images will approximate the structural ODS's.

# 3 Experimental Test Setup for the Wind Turbine Blade (WTB)

In order to extract the dynamic behavior of the baseline and damaged WTB, the test bed shown in Figure 4 was used. A Skystream® 4.7$^{TM}$ WTB was clamped at the root and excited by a modal hammer providing a force impulse. The sequence of images/video of the vibrating WTB was captured using a single 4-megapixel (2048×2048 CMOS sensor) PHOTRON® high-speed camera equipped with a 14-mm lens, as shown in Figure 4(b). The dynamic range of the camera pixels are 8 bits and for each pixel an intensity value in the interval [0-255] is assigned. The frame rate of the high-speed camera was set up to record the sequence of images at 500 frames per second (fps). The sequence of images of the vibrating wind turbine blade is measured up to 4 seconds (2000 frames).

Both the baseline and damaged WTBs are excited by impulses applied approximately at 43cm from the clamped root of the WTB. Moreover, for both the baseline and damaged WTBs, the excitation amplitude and duration are kept approximately consistent between the different tests. The high-speed camera was placed above the WTB, and the approximate distance between the camera and the WTB was 1.8m. This configuration enabled the high-speed camera to capture the full in-plane bending motion of the WTB (i.e. flapwise direction).

Figure 4(c) shows enhanced image data to demonstrate the field of view for the high-speed camera including both the tip and the root. For better visualization in the rest of the article, contrast will be enhanced to better illustrate the testing results.

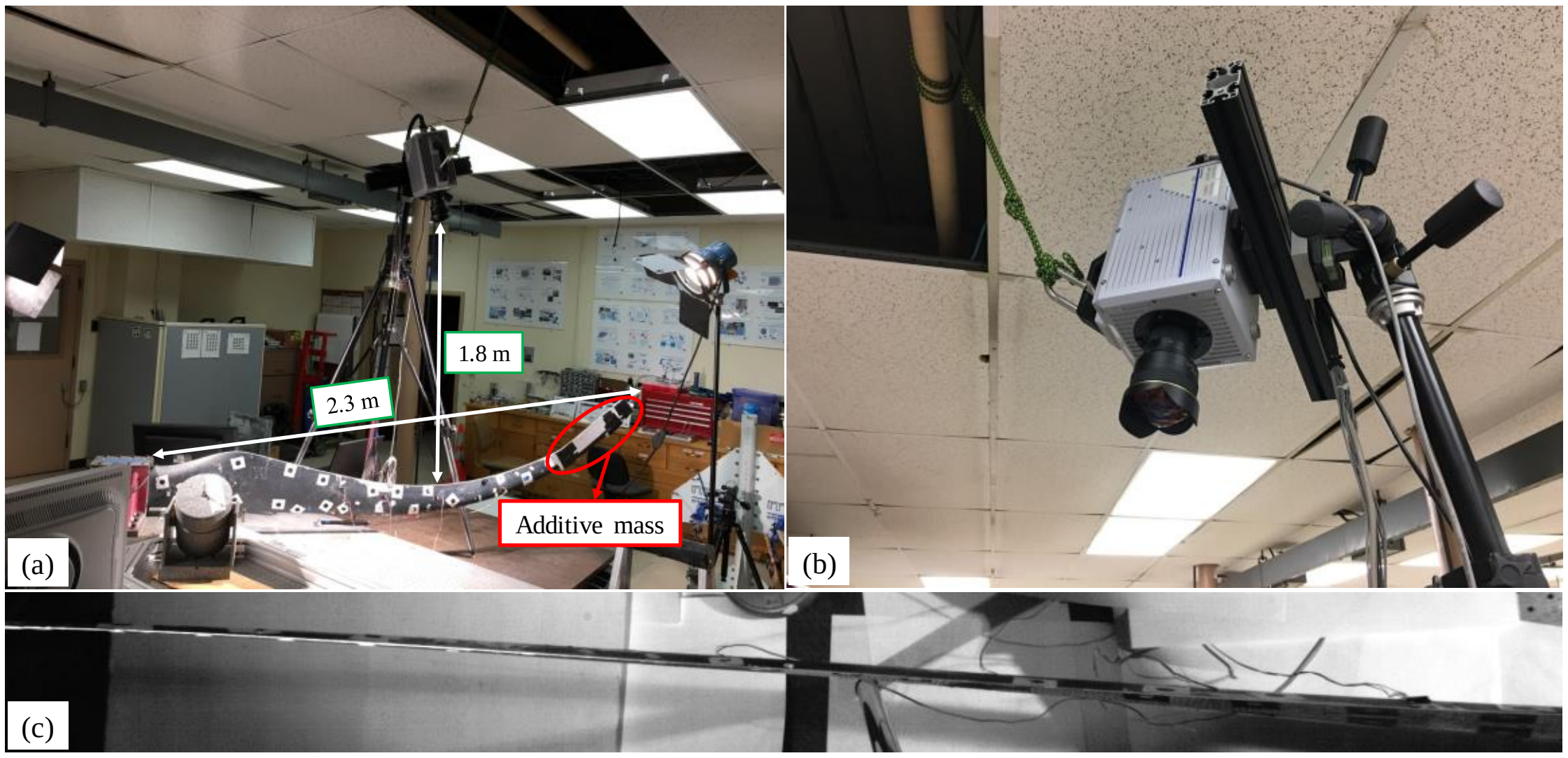

Figure 4. (a) 2.3-m Skystrem® 4.7 cantelivered WTB and the additive mass representing induced damage; (b) Four megapixels PHOTRON high-speed camera equipped with 14-mm lens; (c) Enhanced image data showing the field of view of the high-speed camera

In order to induce temporary damage to the WTB without making it unusable for future testing, additive mass (0.385 kg) was mounted at the tip, which is about 5% of the total WTB mass. The additive mass emulates damage analogous to an icing condition on WTB where frozen water adds unwanted mass to the structure and could potentially decrease the efficiency of the WTB and hinder the functionality while operating. After acquiring vibrational video data from the baseline and damaged WTB, PME and motion magnification are applied to extract the dynamic characteristics of the structures. Later in this paper, the changes in resonant frequencies and operating deflection shapes are used for decision-making regarding the health of the structure.

# 4 Results and Discussion

In Sections 2 and 3, the theoretical background on PME and the experimental test-bed used for the case studies are introduced. This section primarily focuses on the post-processing of the recorded sequence of images from the baseline and damaged WTB in order to estimate the resonant frequencies and operational deflection shapes which are being used later to detect the occurrence of damages.

Figure 5 shows the overall workflow performed within this paper to extract the modal parameters for either the baseline or damaged WTB. First, over a duration of 4 second, 2000 images are captured by the high speed camera during which the blade was impacted by a hammer near its root to induce a subtle vibration. Second, PME is applied to extract the motion information from the acquired images. The extracted motion is then transferred to frequency domain, and the resonant frequencies of the structure are obtained using a peak-picking procedure. The extracted resonant frequencies will be used as the center frequencies in the motion magnification step. In the next step, in order to obtain a better visualization in presenting the results in the paper, the quality of images has been increased by changing the dynamic ranges of pixel intensities to resolve the poor illumination condition. This low contrast issue is common while working with high-speed cameras, because the shutter time for high-speed cameras is very short in order to sample with high frame rates. This reduces the exposure of the sensors to light, and therefore causes contrast degradation in images. It should be emphasized that this enhancement is only for visualization in presenting the extracted operational deflection shapes, but not required by the PME and magnification algorithms. However, using the full the dynamic range of the pixel intensities is always beneficial for computer vision algorithms. In the third step, the center frequencies for motion magnification are selected, according to the resonant frequencies extracted in the PME stage. The resulting videos reveal the operating deflection shape of the WTB at the resonant frequencies of interest and will be similar to the mode shapes, as long as there is sufficient mode separation. In the last step, the extracted operating deflection shapes are quantified by means of an edge detection on the motion-magnified videos. The detailed discussion regarding each of the steps is presented in the following sections.

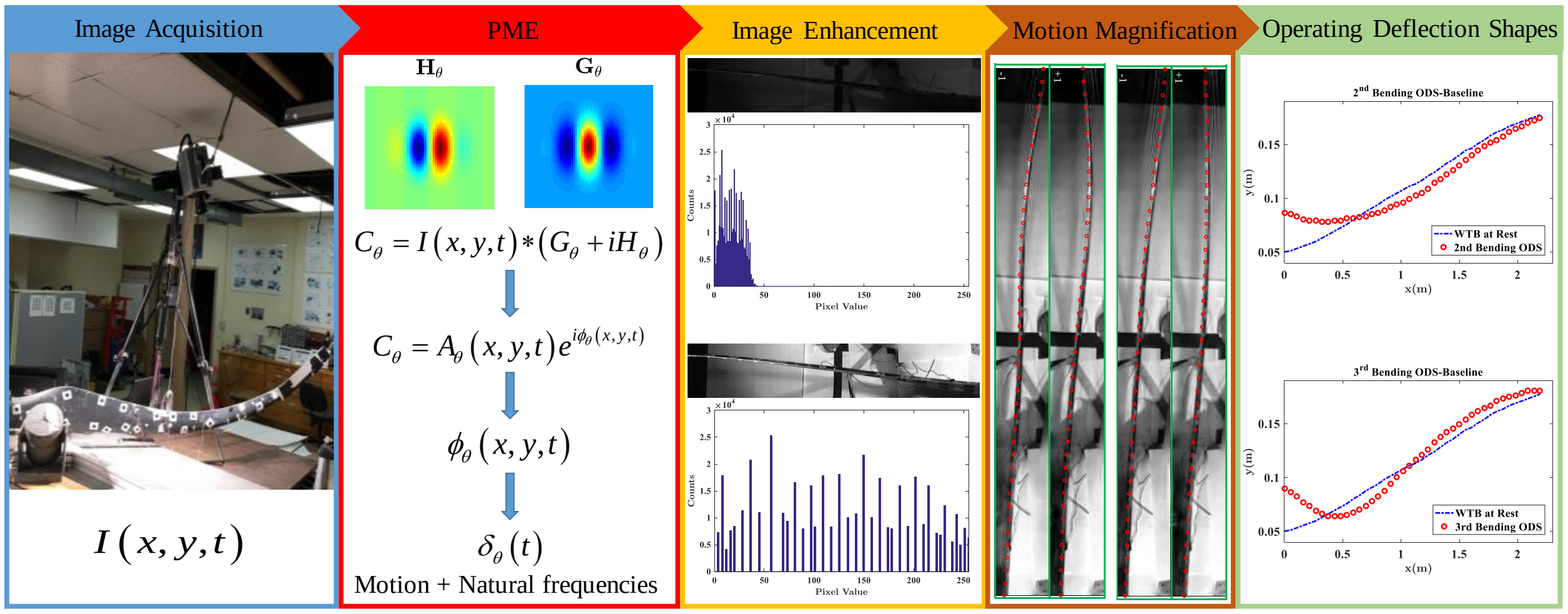

Figure 5. Workflow diagram of operating deflection shapes extraction with small vibrational response via PME

## 4.1 Resonant Frequencies

As a benchmark, the resonant frequencies of the WTB are firstly estimated by performing EMA using accelerometers [40], and the frequencies and corresponding shapes are shown in Figure 6. The extracted frequencies via EMA are used to evaluate the performance of the proposed PME approach of the baseline WTB without damages introduced. EMA test is performed using 13 tri-axial PCB® accelerometers (model number: 356A32, nominal sensitivity: 100 mv/g) are used to measure the vibrational response of the structure due to the excitation from a PCB® modal hammer (model number 086C03). Each of the accelerometers weights 5.4 g, and the total additive mass to the wind turbine blade from the accelerometers is about 70 grams. A twenty-channel LMS data acquisition system is used to record the accelerations and the input force to the wind turbine blade. Moreover, all the EMA test equipment are removed from the wind turbine blade before starting the optical measurement test. The modal parameters are estimated using PolyMAX curve fitting technique .The frequencies obtained from conventional acceleration-based EMA and the PME are compared in Figure 7(a), which indicates a very good agreement in determining frequencies from the sequence of images. The error between the resonant frequencies provided by the proposed non-contact method and the conventional EMA data is less than 0.6 Hz for all of the ODS's.

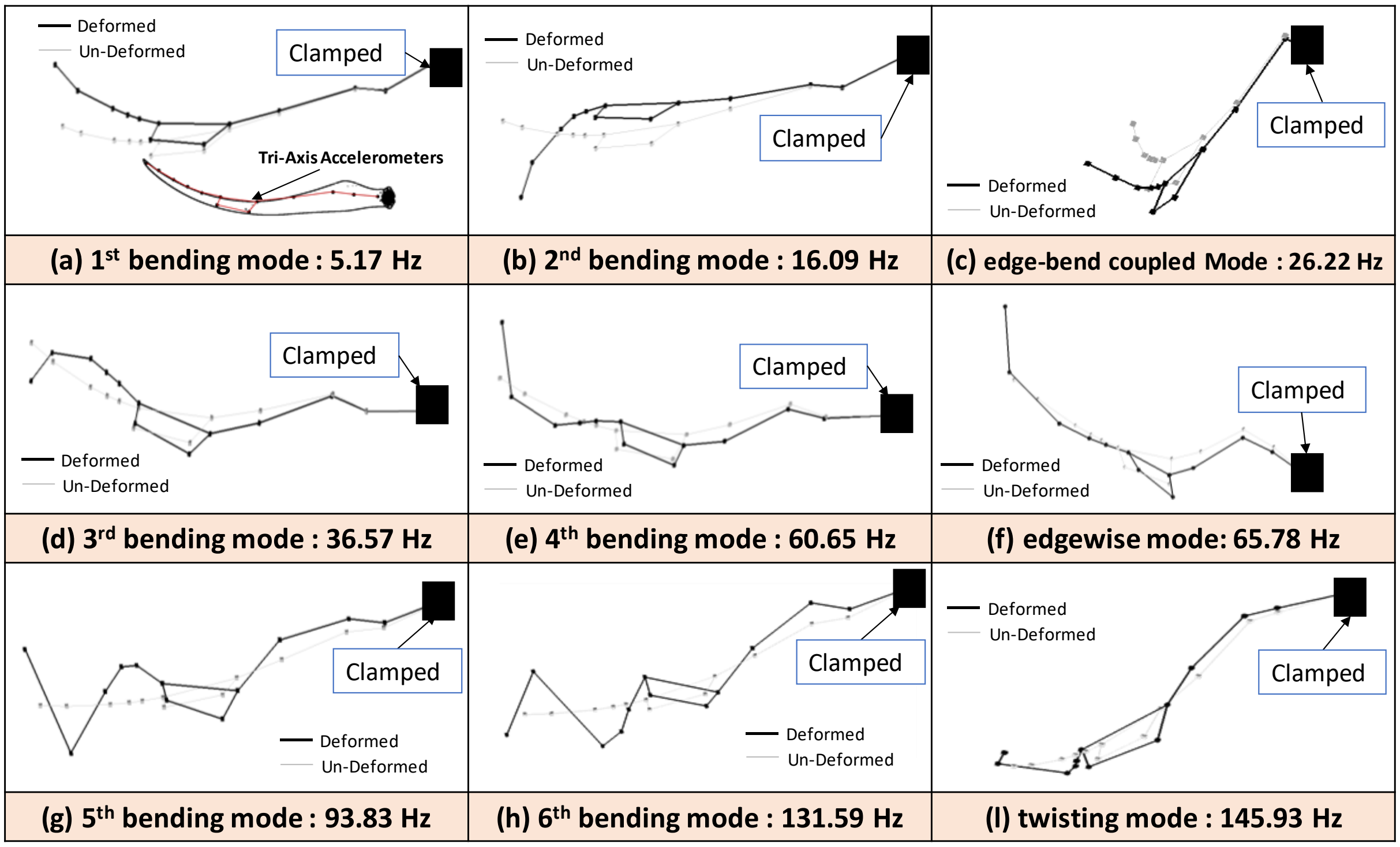

Figure 6. The first 9 mode shapes of the cantilevered wind turbine blade with their corresponding resonant frequencies [40, 41]

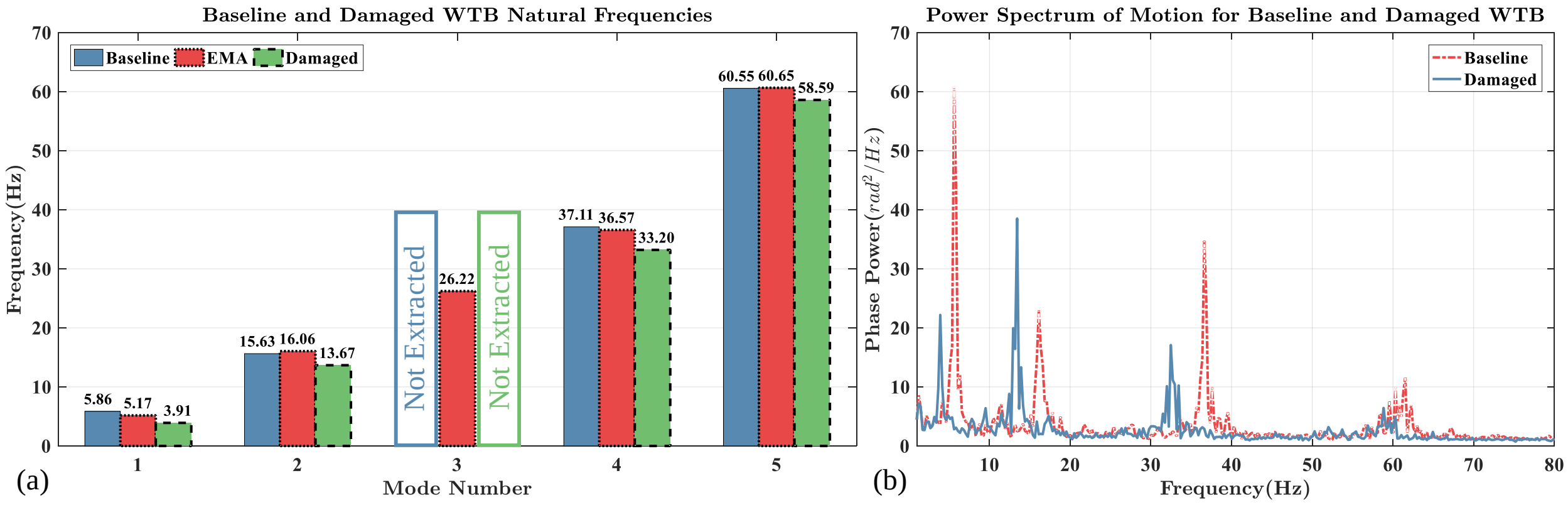

Figure 7. (a) Resonant frequencies of the baseline WTB using PME and EMA in comparision with the resonant frequencies of the damaged WTB; (b) Frequency domain respesentation of estimated motions of baseline and damaged WTB, measured at the blade tip

Since the single camera measures 2D information without depth information, PME is limited to extract planar motions parallel to the camera's image plane (i.e. flapwise direction). The third and sixth modes (26.22 Hz and 65.78 Hz) are the edgewise modes of the WTB which are normal to the camera's image plane in this study. Therefore, these operating deflection shapes cannot be detected by means of using PME on the captured sequence of images. Except for the edgewise bending modes, all the other resonant frequencies are in good agreement with the results compared to the traditional EMA.

Since the estimated resonant frequencies from PME have been validated based on the EMA results, the motion estimation algorithm can be applied on damaged WTB to extract its resonant frequencies with high confidence as well. Figure 8(b) shows the frequency content of the estimated motion for baseline and damaged WTB. As expected adding mass to the WTB in order to simulate the icing condition, makes the resonant frequencies of the damaged WTB to decrease. Considering the shift in the resonant frequencies as shown in Figure 7, the presence of damage can be easily detected by using PME via non-contact measurements without the need for conventional accelerometers. In this study we use simple comparative inference from the features, and for further investigation, statistical and probabilistic data analysis can be performed which is not the scope of this context.

## 4.2 Operating Deflection Shapes (ODS's)

For some structures, changes in operating deflection shapes may also be used to reveal damage presence apart from identifying changes in the resonant frequencies. Changes in ODS's (as an extra damage sensitive feature) may increase the reliability of the decision-making regarding the health status of structure. Moreover, the ODS's preserves spatial information, so damage localization is also possible for future SHM activities.

As mentioned earlier, one of the challenges in image acquisition acquired by high-speed cameras is the need for sufficient illumination and contrast, due to the high frame rate and resulting short shutter time. To address this issue, before applying the motion magnification algorithm to extract the operating deflection shapes, the visibility is increased by means of image enhancement. As shown in Figure 8(a), the original image brightness and resulting contrast of the WTB is too dark and the motion magnified video reconstructed from this set of images will not be visually informative. Increasing the contrast of images via manipulation of the distribution of pixel intensities is a practical approach to overcome this problem. Figure 8(b) shows the histogram of the pixel intensities for the original dark image, and obviously, the pixel intensities are concentrated in a narrow band and lead to low contrast and visibility. By resetting the dynamic range of the pixel intensities, they are mapped onto a wider interval, and lead to a better presentation as shown in Figure 8(c) and the updated histogram in Figure 8(d). This action only increases the contrast and visibility of original image without effecting any of the motion information. Moreover, using the full dynamic range of pixel intensities can help PME to provide better estimations of the motion. In the original image the pixel intensities are limited to the interval [0 50] while after image enhancement the full 8 bit range of pixels will be taken into account [0 255].

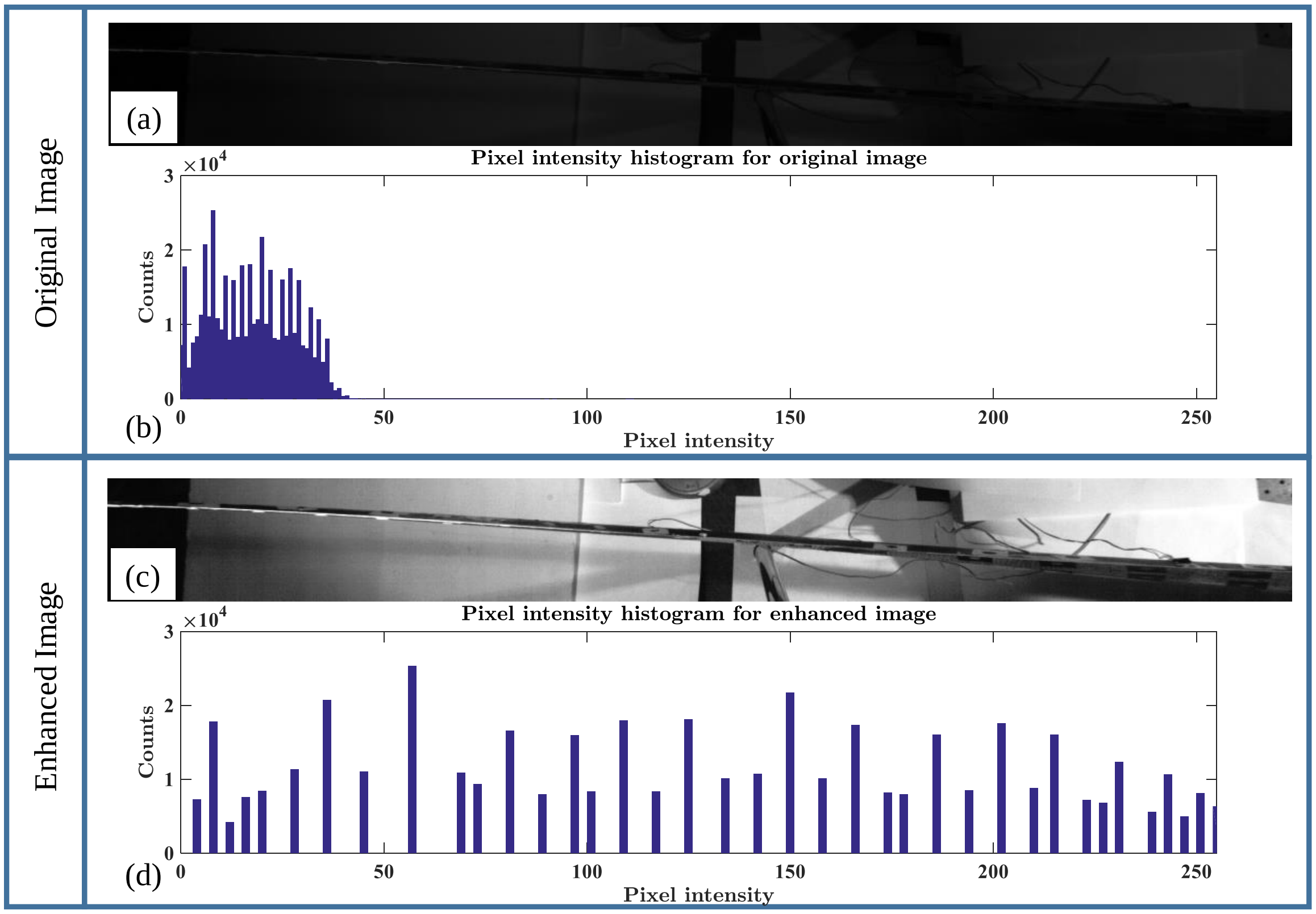

Figure 8. (a) Original image; (b) Pixel intensity histogram for the original image; the maximum intensity is 255; (c) Enhanced image; (d) Pixel intensity histogram for enhanced image

In order to extract the operational deflection shapes from the sequence of images, a center frequency ( $f_c$ ) for motion magnification needs to be selected, which is the corresponding resonant frequency of the structure as estimated in previous section through PME. Then the width of the pass band of the motion magnification filter ( $b$ ) is tuned to only contain that specific frequency content for that particular resonance. Applying the band-pass filter with above-mentioned parameters will reconstruct a video, which merely contains the operating deflection shape of the structure at that desired resonant frequency. In Figures 9 to 12, snap shots of the videos with magnified motions for the first four bending operating deflection shapes are shown. Each pair of the snap shots are selected at the frames with maximum deflection, which correspond to a $180^\circ$ phase difference denoted as “+1” and “-1” in the associated figures.

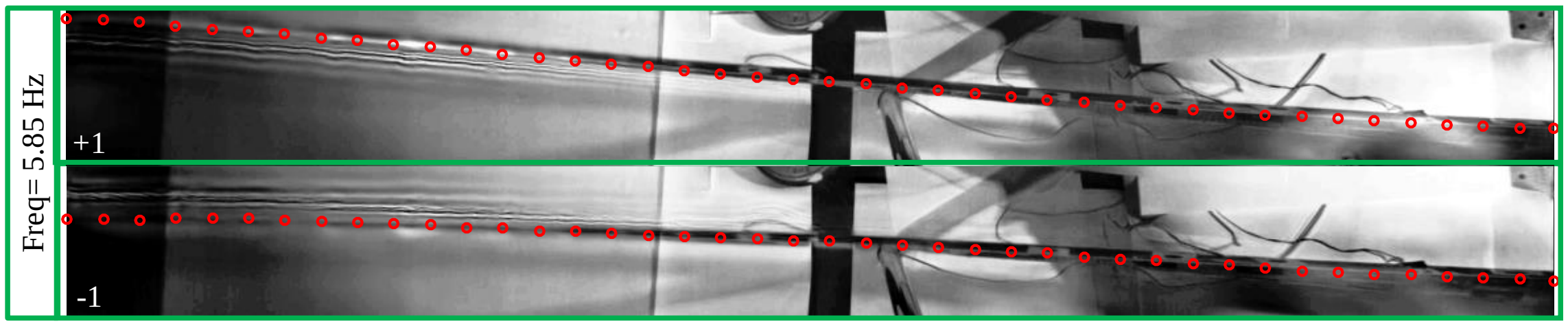

Figure 9. Snap shots of the motion magnified video for the baseline WTB at its maximum deflection. Motion magnification parameters: $f_c = 5.85\text{Hz},\ b = 3\text{Hz},\ \alpha = 25$

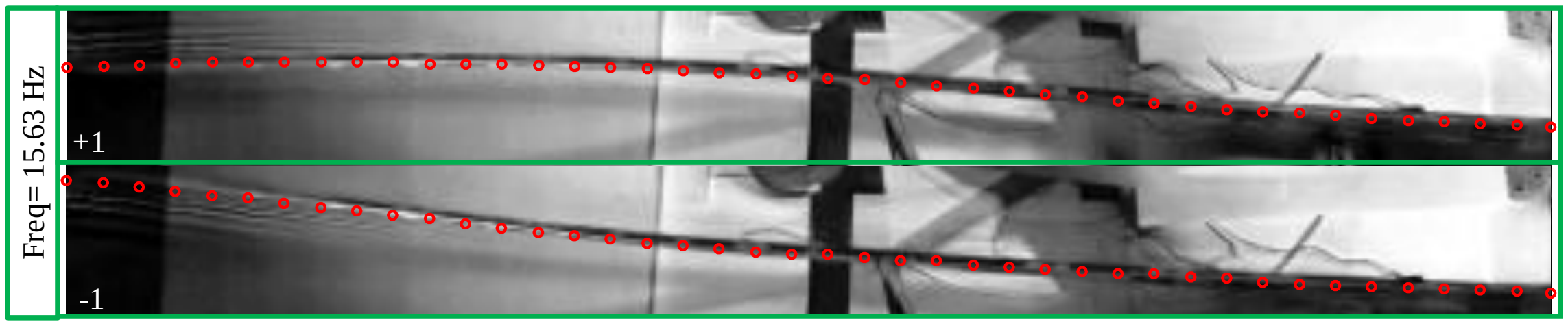

Figure 10. Snap shots of the motion magnified video for the baseline WTB at its maximum deflection. Motion magnification parameters: $f_c = 15.63\text{Hz},\ b = 3\text{Hz},\ \alpha = 35$.

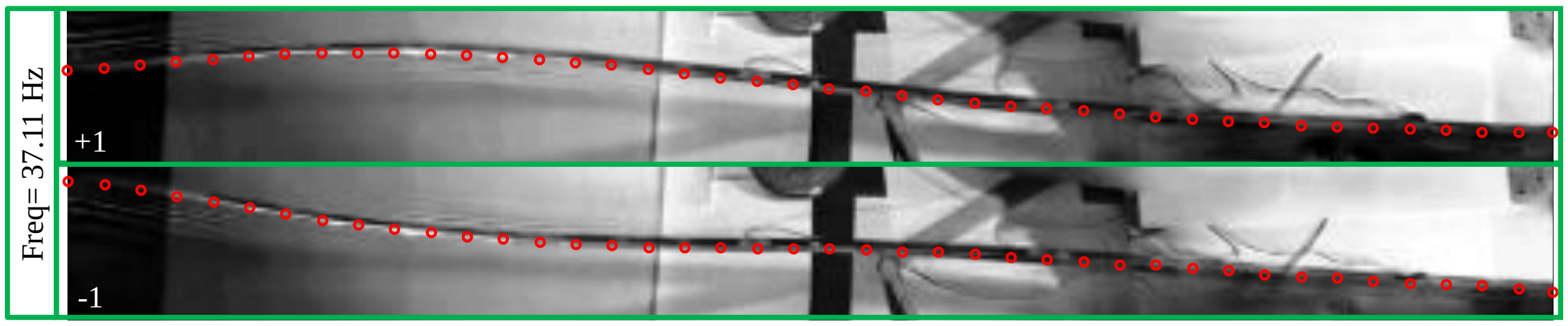

Figure 11. Snap shots of the motion magnified video for the baseline WTB at its maximum deflection. Motion magnification parameters: $f_c = 37.11\text{Hz},\ b = 3\text{Hz},\ \alpha = 75$.

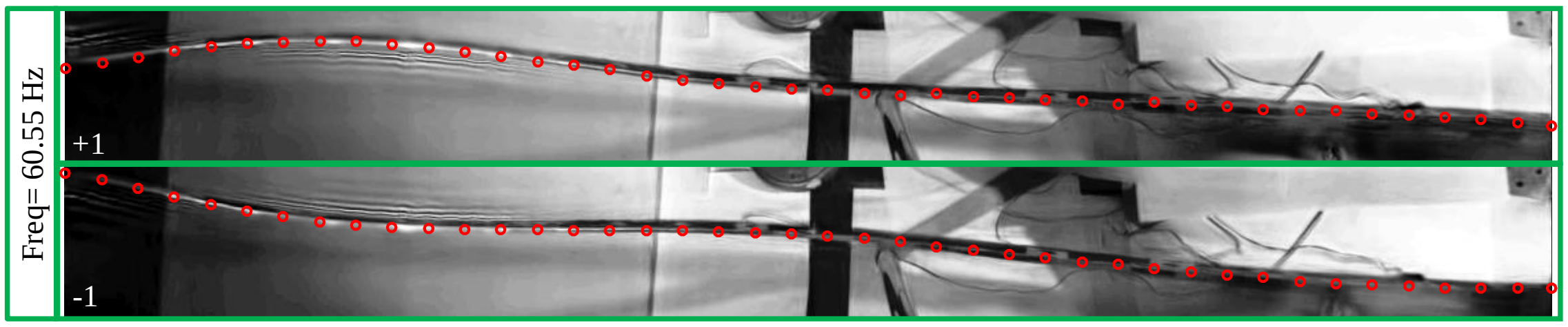

Figure 12. Snap shots of the motion magnified video for the baseline WTB at its maximum deflection. Motion magnification parameters: $f_c = 60.55\text{Hz},\ b = 3\text{Hz},\ \alpha = 150$.

In this demonstration, higher magnification factors ( $\alpha$ ) are selected for higher order modes, since higher order modes have less displacement contribution to the response and the overall magnitude of the motion is smaller. The width of the pass band of the filter in this validation is selected as 3 Hz in all the motion-magnified videos. Applying such narrow-band filters on the original sequence of images will lead to amplified motions, and because of the narrow bandwidth and mode separation, it is unlikely to have other modes contributing significantly in this frequency range. Therefore, the magnified motion may be regarded as the operating deflection shape at the selected resonant frequency. In addition to visually observing the operating deflection shapes, PME is adopted once more on the motion-magnified videos to extract quantitatively the amplified motions. This procedure can grantee that the motion in the magnified video is actually dominated by the desired resonant frequency, and the processed video represents the ODS associated with the selected resonant frequency. As shown in Figure 13, the dotted straight lines represent the selected frequency band applied for motion magnification. After magnifying the motion in the corresponding selected frequency bands, only the motion content at that resonant frequency is significantly amplified.

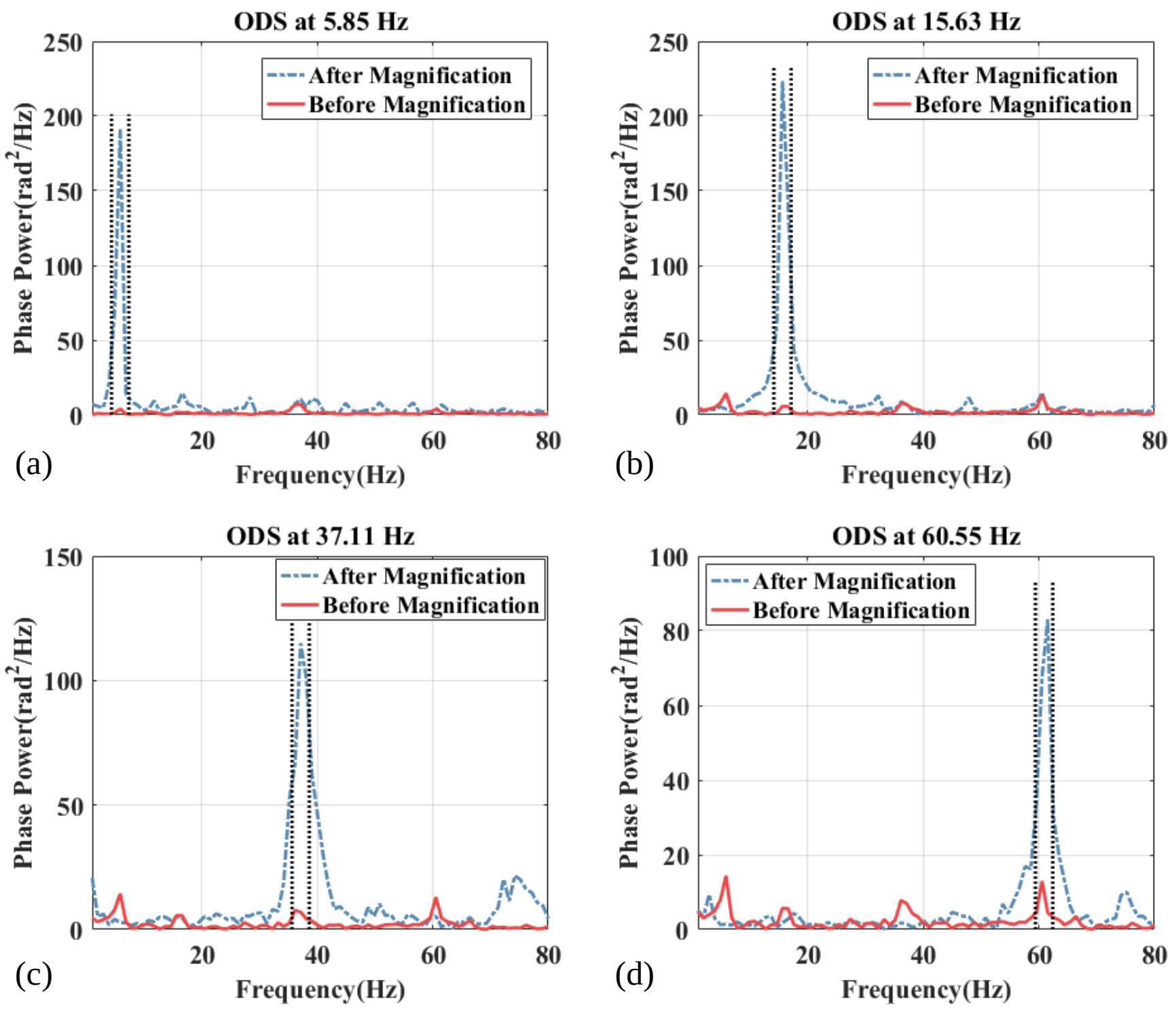

Figure 13. Frequency content of the motion before and after applying motion magnification for the baseline WTB of the video representing: (a) 1st ODS at 5.85 Hz, (b) 2nd ODS at 15.63 Hz, (c) 3rd ODS at 37.11 Hz, (d) 4th ODS at 60.55 Hz

Having quantitative representation of operating deflection shape vectors will enable SHM decision-making. As a widely-adopted technique, the edge detection algorithm will facilitate quantifying the magnified motions [35]. The edge detection algorithm is normally applied to provide the coordinates of pixels associated with the boundary of the object that is being investigated, and in this case, the WTB. However, it is very difficult and time consuming in reality to tune the parameters of the edge detection algorithm to only extract the pixels corresponding to the object of interest. Therefore, the outcomes of the edge detection practice are often cleaned up manually to exclude the irrelevant edges present in the scene, such as sensor wires and background discontinuities that are specifically found in this WTB vibration test. Moreover, at a few points, the edge detection algorithm cannot extract coordinates accurately, and under this circumstance, the location of those specific points have to be corrected manually (based on the other nearby points) to obtain a smooth and more reasonable operating deflection shape. The human interaction mainly concerns the detection of unwanted edges in the background in the field of view. Having sharp edges in the background may cause the edge detection algorithm to produce much noisy outcome and wrong sketch of the moving object, which should be eliminated manually. This procedure includes excluding the outlier points from our dataset. For future work, more automated techniques may be investigated. Speaking of this, quantitative operating deflection shape extraction from the motion-magnified videos is not currently completely automatic and supervision is necessary to obtain a reliable outcome. After applying edge detection, Figure14 shows the un-deformed WTB overlaid on the extracted deflected shapes (red circles) for the first four operating bending deflection shapes. The pixel locations are calibrated based on the length of the WTB, which is 2.3m. The form of the extracted operating deflection shapes including the location of the nodes and anti-nodes are in good agreement with modes extracted from EMA. It should be also highlighted that the WTB at its rest position in the selected field of view for the camera (Figures 9-12 and 16-19) has a tilted orientation; therefore, the extracted ODS's in Figures 14, 20 and 21 are tilted as well. The difference between the orientations in the extracted ODS's and the camera views are due to the aspect ratio of the subplots – the *y*-scale in Figure 14, 21 and 22 are stretched for a clearer visualization.

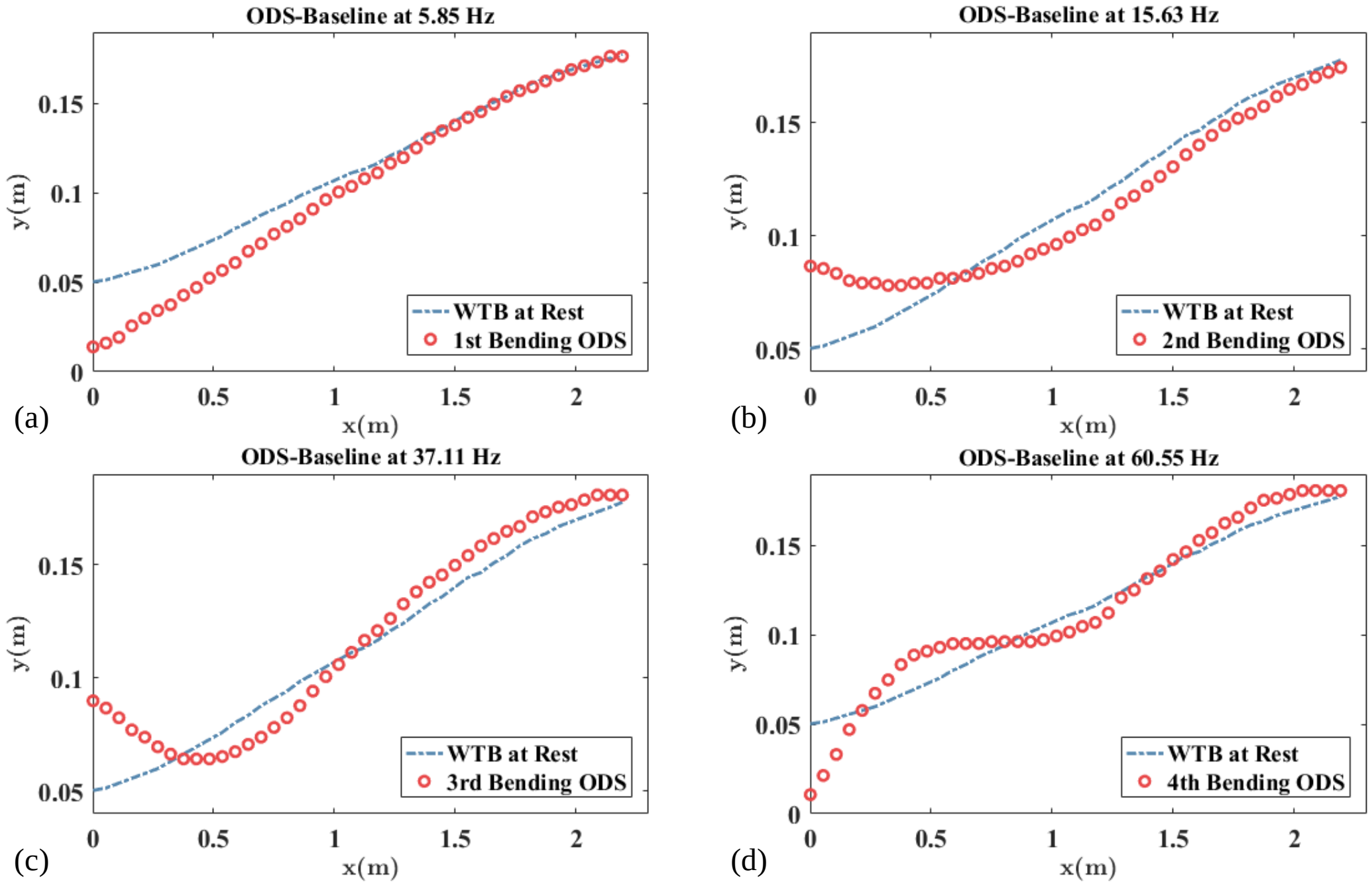

Figure 14. ODS's of the baseline WTB: (a) 1st ODS at 5.85 Hz, (b) 2nd ODS at 15.63 Hz, (c) 3rd ODS at 37.11Hz, (d) 4th ODS at 60.55 Hz

In order to evaluate the accuracy of the extracted ODS the MAC value between the acquired ODS's from PME and EMA data is computed. MAC is evaluated at the location of the accelerometers and as shown in figure 15 a high correlation (higher than 85%) is achieved between the ODS's obtained from PME and EMA which validates the results provided by PME.

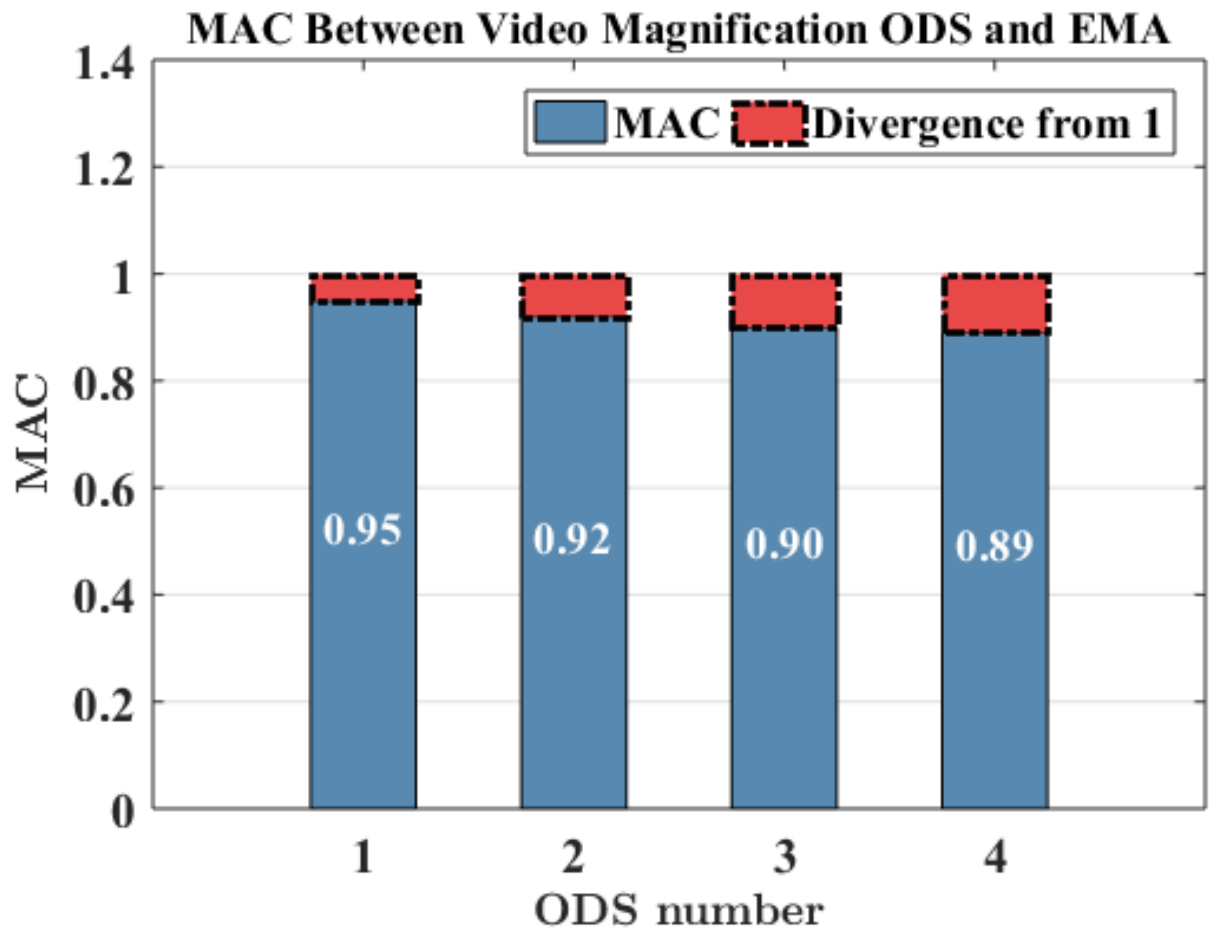

Figure 15. MAC evaluation of extracted ODS's using PME with EMA data as the benchmark

The same procedure may be applied on the damaged WTB. Initially, the contrast of the sequence of images are increased. Then the motion magnification algorithm is applied on the enhanced video within a narrow frequency band about the resonant frequencies. The resulting videos are dominated by the operating deflection shape of the damaged WTB at each of the modes of interest. Similarly, Figures 16 to 19 show two snaps shots of each of the operating deflection shapes at maximum deflections, and the frequency domain representation of the magnified displacement response at the tip of the damaged WTB is shown in Figure 20. Afterwards, the quantitative operating deflection shapes are extracted for the damaged WTB by using the edge detection algorithm. The first four bending operating deflection shapes of the damaged structure are shown in Figure 21.

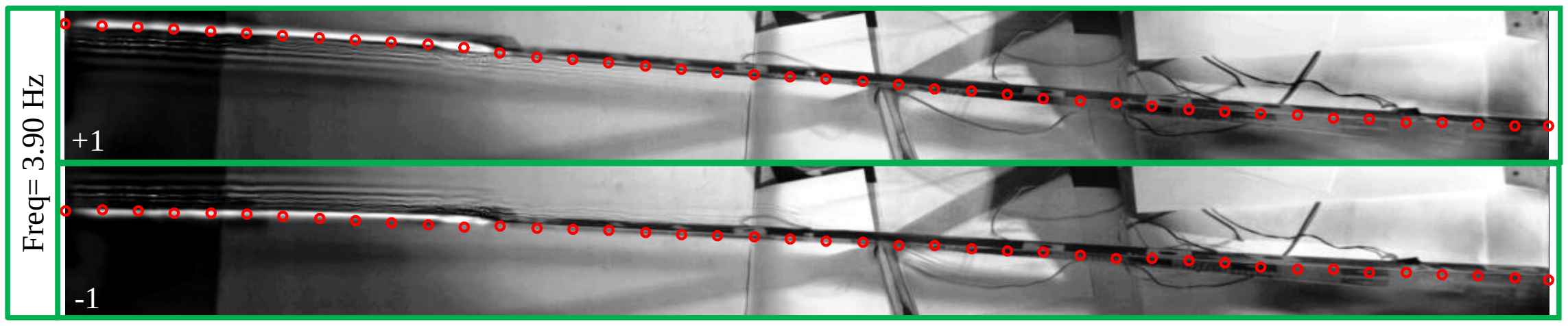

Figure 16. Snap shots of the motion magnified video for the damaged WTB at its maximum deflection. Motion magnification parameters: $f_c = 3.90\text{Hz}, b = 3\text{Hz}, \alpha = 25$.

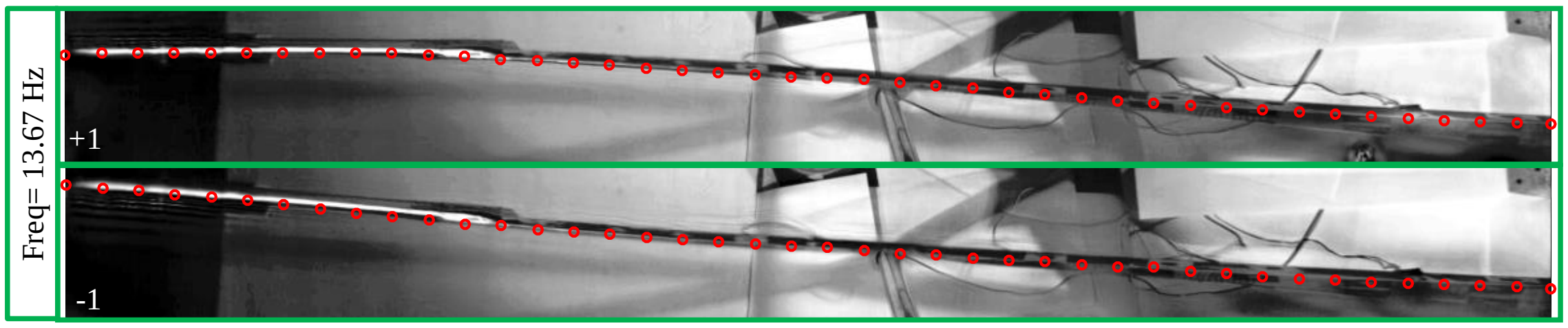

Figure 17. Snap shots of the motion magnified video for the damaged WTB at its maximum deflection. Motion magnification parameters: $f_c = 13.67\text{Hz}, b = 3\text{Hz}, \alpha = 35$.

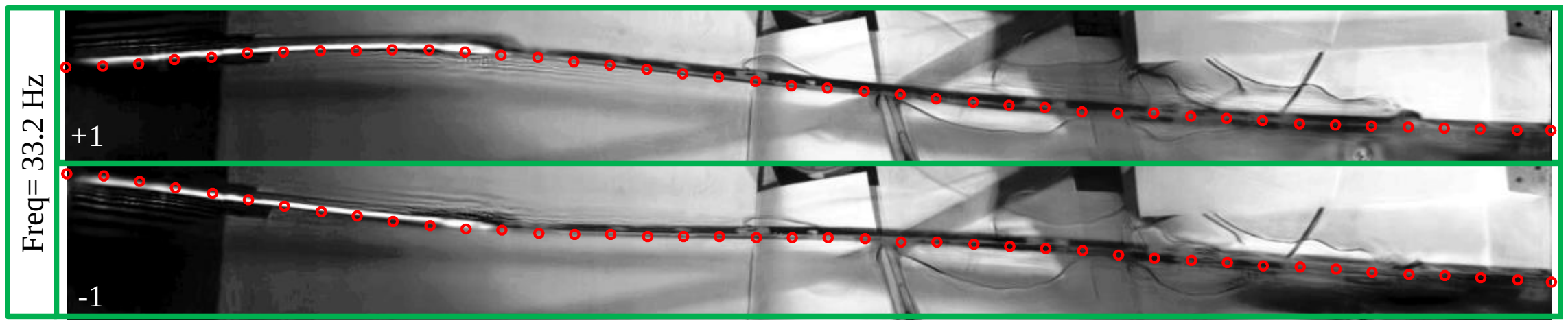

Figure 18. Snap shots of the motion magnified video for the damaged WTB at its maximum deflection. Motion magnification parameters: $f_c = 33.2\text{Hz}, b = 3\text{Hz}, \alpha = 75$.

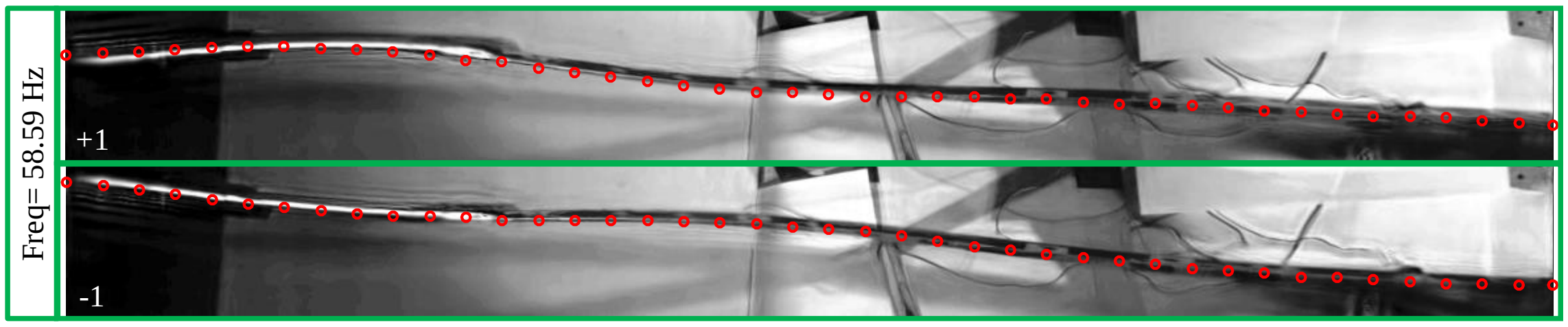

Figure 19. Snap shots of the motion magnified video for the damaged WTB at its maximum deflection. Motion magnification parameters: $f_c = 58.59\text{Hz}, b = 3\text{Hz}, \alpha = 150$.

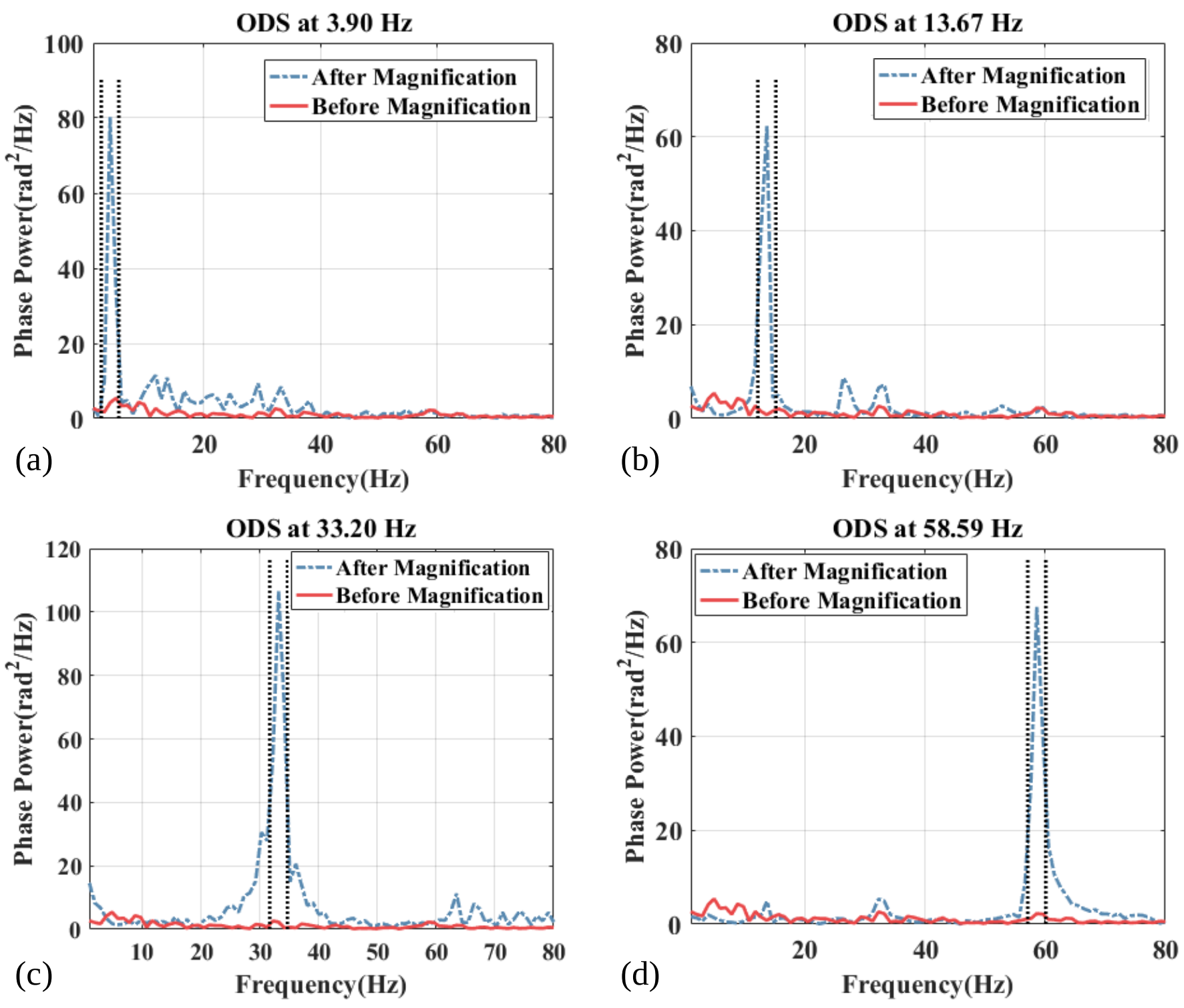

Figure 20. Frequency content of the motion before and after applying motion magnification for the damaged WTB for the video representing: (a) 1$^{st}$ ODS at 3.90 Hz, (b) 2$^{nd}$ ODS at 13.67 Hz, (c) 3$^{rd}$ ODS at 33.20 Hz, (d) 4$^{th}$ ODS at 58.59 Hz

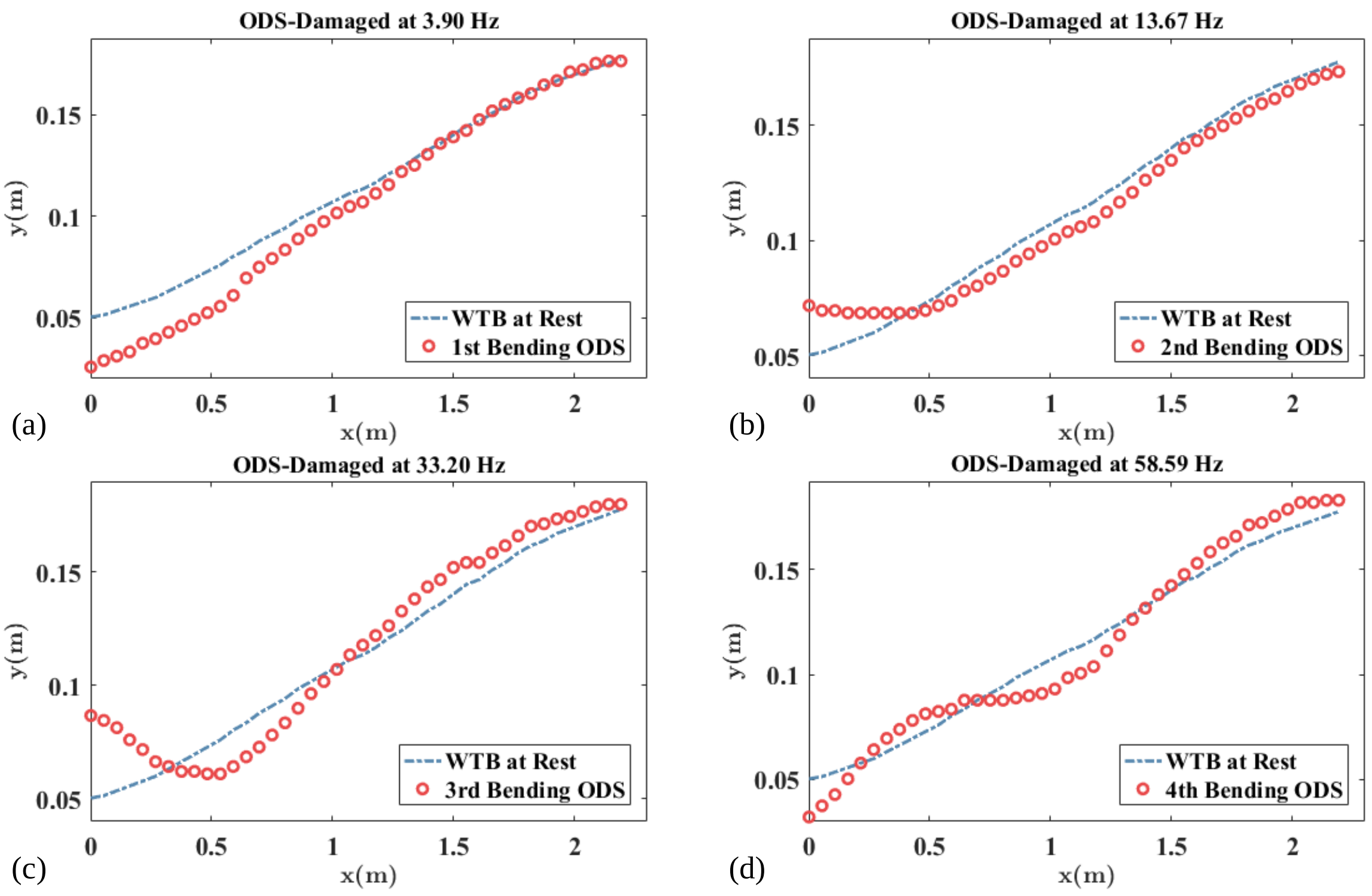

Figure 21. ODS's of the damaged WTB: (a) 1$^{st}$ ODS at 3.90 Hz, (b) 2$^{nd}$ ODS at 13.67 Hz, (c) 3$^{rd}$ ODS at 33.20 Hz, (d) 4$^{th}$ ODS at 58.59 Hz

Figure 22 shows the quantified operating deflection shapes of baseline and damaged WTB. Although the deviations of the operating deflection shapes for the damaged case are not obvious for the first and third bending ODS, the second and fourth operating deflection shapes show clear changes in WTB bending ODS form, as well as having differences in the locations of the nodes and anti-nodes that indicate a presence of damage. Since the vibrations of the wind turbine blade are small, the analysis is taking place is the linear elastic region. Therefore, the ODS of the wind turbine blade will be proportional to the amplitude of the impact force for a set of impacts at the same location. Moreover, the location of nodes and anti-nodes are also independent of the amplitude of the impact force. To quantitatively compare the operating shapes between the baseline and damaged cases, the Modal Assurance Criterion (MAC) is used for each ODS as shown in Figure 23. A MAC value of unity indicates complete shape similarity while a value of zero indicates that there is no similarity between the operating deflection shapes. Since the consistent locations of impact are selected for both the baseline and the damaged cases the deviations in MAC value can be traced back to the changes in the dynamics of the structure due to the damage (additive mass). The similarity between the operating deflection shapes decreases considerably in bending ODS's two and four as expected, which indicates the presence of damage in the WTB.

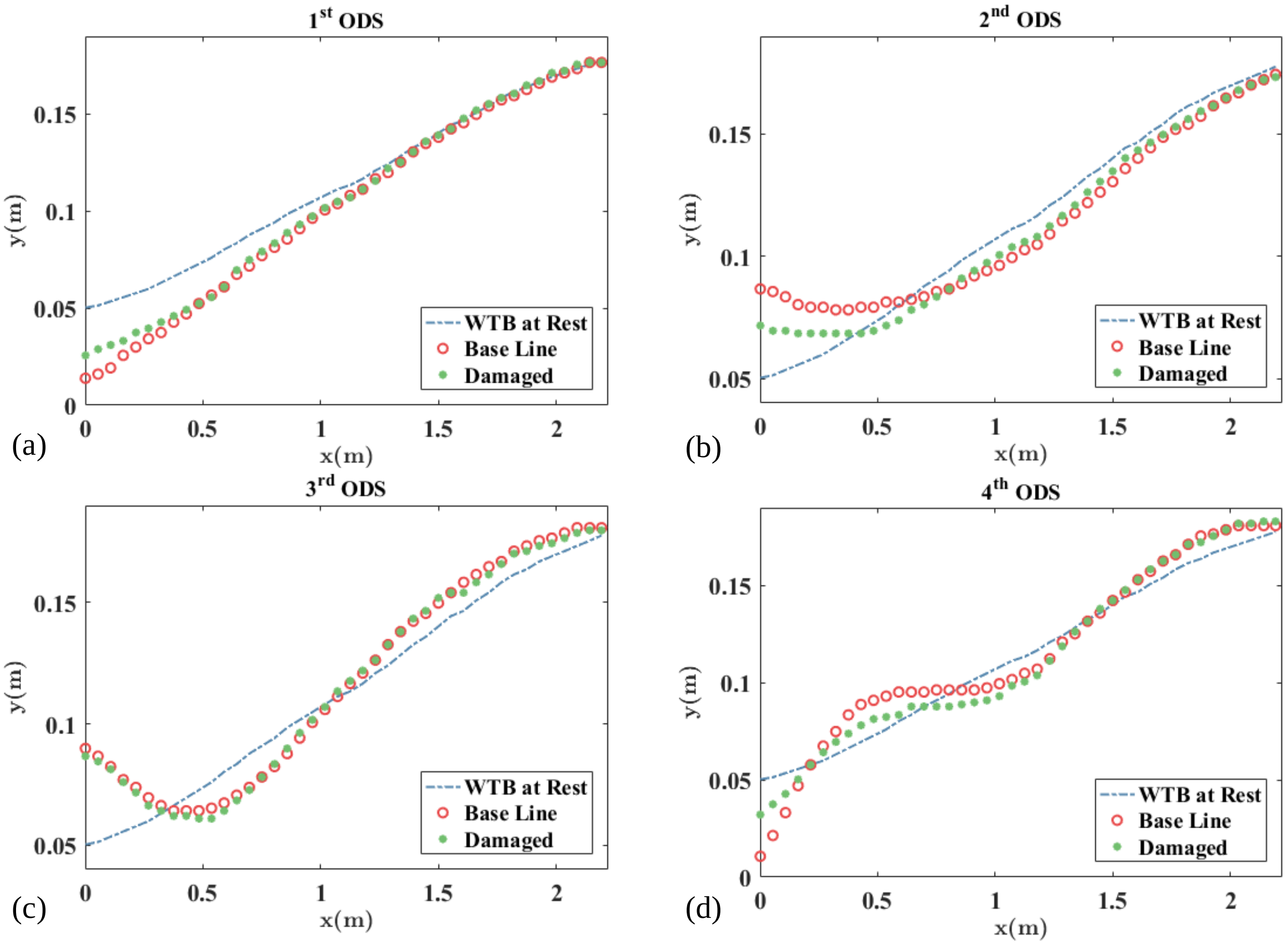

Figure 22. Overlaid plot of the ODS's of baseline and damaged WTB: (a)1st ODS, (b)2nd ODS, (c)3rd ODS, (d)4th ODS

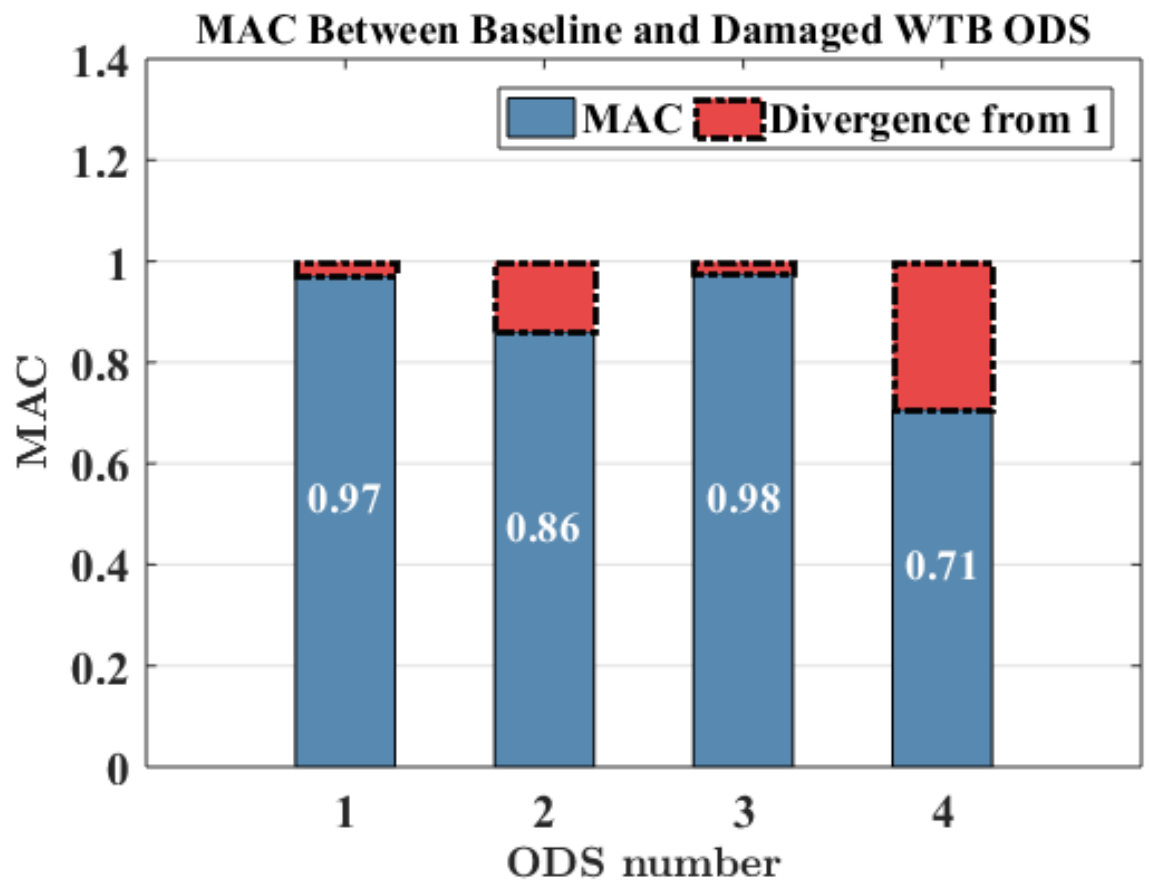

Figure 23. MAC value between the oprating deflection shapes vectors for baseline and damaged WTB

Based on the results presented in Figures 22 and 23, PME is able to detect the presence of damage in the WTB and is likely able to be used for damage detection or SHM for other types of structures and applications.

# 5 Conclusion

This paper proposes and demonstrates Phase-based Motion Estimation (PME) and a motion magnification algorithm to perform non-contact structural damage detection of a wind turbine blade. The paper is the first effort formulating the PME to 2-dimesinal scenarios, and applying PME and motion-magnification on an industrial and complex geometry structure. The novel approach can be applied to numerous structures and has the benefit of not inducing any mass-loading or affecting the original structural stiffness, does not require instrumented sensors or a surface treatment (e.g. speckle pattern), and can be achieved by using a single commercially available camera. Image enhancement is also conducted before actually extracting operating deflection shapes in the proposed work, by means of manipulating the histogram of pixel intensities in the acquired videos, and the quality of each image frame is dramatically improved even when the high-speed camera acquired images suffer from limited lighting illumination and contrast. Compared to other conventional non-contact image sensing techniques, such as 3D DIC, 3DPT that can be used for SHM, the adopted phase-based approach has the advantages of avoiding surface preparation and the application of a speckle pattern on the structure by only using the natural features of the surface itself. Moreover, the computation consumption of this approach is much less than the other image sensing techniques because the data from one camera is being processed. Additionally, only a single camera sensor is required to obtain structural dynamic information such as resonant frequencies and operating deflection shapes. However, because of the 2D nature of the image plane, three-dimensional motion by using a single camera sensor is not readily achievable. A damage detection effort on a 2.3m commercially available WTB was conducted, in which a damage was introduced to the WTB by adding extra mass to the tip of the blade. PME was used to extract the dynamic motion information and the resonant frequencies of the structure from the acquired video. Changes in the structural resonant frequencies are present when the changes in structural mass are made. The operating deflection shapes of the baseline and damaged structures are estimated via the phase-based motion magnification and edge detection algorithm and are quantified by using the modal assurance criterion. The comparison between the two states reveals that the proposed optically based approach can be used for damage detection and to identify the system dynamic characteristics, including resonant frequencies and operating deflection shapes. Moreover, the magnified motions in the manipulated videos provide a better visualization of the higher order structural modal characteristics. However, extracting the quantitative ODS from motion-magnified videos and also selecting the magnification factors require human interaction, therefore the method is not fully automated at this time. Additional studies are needed to improve and automate the edge detection in a more robust way, in order to eliminate the human interaction in the analysis process. Other than that, the parameter determination in designing the optimal filter characteristics for video motion magnification is another issue, which needs to be addressed, in terms of the center frequency and bandwidth, especially when the structure is geometrically complex and the structural modes are not well separated in the frequency domain. Moreover, for future studies, sensitivity analysis on damage detection using PME is absolutely necessary. Other than additive mass, other types of damages should be also studied in order to evaluate the full potential of the PME and motion magnification for SHM applications.

# 6 References

[1] A. Deraemaeker, E. Reynders, G. De Roeck, J. Kullaa, Vibration-based structural health monitoring using output-only measurements under changing environment, Mechanical systems and signal processing, 22 (2008) 34-56.
[2] C.R. Farrar, K. Worden, An introduction to structural health monitoring, Philosophical Transactions of the Royal Society of London A: Mathematical, Physical and Engineering Sciences, 365 (2007) 303-315.
[3] S.W. Doebling, C.R. Farrar, M.B. Prime, A summary review of vibration-based damage identification methods, Shock and vibration digest, 30 (1998) 91-105.
[4] A. Raghavan, C.E. Cesnik, Review of guided-wave structural health monitoring, Shock and Vibration Digest, 39 (2007) 91-116.
[5] A. Ebrahimkhanlou, B. Dubuc, S. Salamone, Damage localization in metallic plate structures using edge-reflected lamb waves, Smart Materials and Structures, 25 (2016) 085035.
[6] C.U. Grosse, M. Ohtsu, Acoustic emission testing, Springer Science & Business Media, 2008.
[7] A. Ebrahimkhanlou, S. Salamone, Acoustic emission source localization in thin metallic plates: A single-sensor approach based on multimodal edge reflections, Ultrasonics, 78 (2017) 134-145.
[8] E.G. Henneke, K.L. Reifsnider, W.W. Stinchcomb, Thermography—an NDI method for damage detection, JOM, 31 (1979) 11-15.
[9] D. Balageas, C.-P. Fritzen, A. Güemes, Structural health monitoring, John Wiley & Sons, 2010.
[10] C.R. Farrar, K. Worden, Structural health monitoring: a machine learning perspective, John Wiley & Sons, 2012.
[11] D.J. Ewins, Modal testing: theory and practice, Research studies press Letchworth, 1984.
[12] C. Niezrecki, P. Avitabile, J. Chen, J. Sherwood, T. Lundstrom, B. LeBlanc, S. Hughes, M. Desmond, A. Beattie, M. Rumsey, Inspection and monitoring of wind turbine blade-embedded wave defects during fatigue testing, Structural Health Monitoring, (2014) 1475921714532995.
[13] D. Montalvao, N.M.M. Maia, A.M.R. Ribeiro, A review of vibration-based structural health monitoring with special emphasis on composite materials, Shock and Vibration Digest, 38 (2006) 295-326.
[14] M. Ashory, Correction of mass-loading effects of transducers and suspension effects in modal testing, in: SPIE proceedings series, Society of Photo-Optical Instrumentation Engineers, 1998, pp. 815-828.
[15] O. Cakar, K. Sanliturk, Elimination of transducer mass loading effects from frequency response functions, Mechanical Systems and Signal Processing, 19 (2005) 87-104.
[16] P. Poozesh, J. Baqersad, C. Niezrecki, P. Avitabile, E. Harvey, R. Yarala, Large-area photogrammetry based testing of wind turbine blades, Mechanical Systems and Signal Processing, (2016).
[17] W. Fan, P. Qiao, Vibration-based damage identification methods: a review and comparative study, Structural Health Monitoring, 10 (2011) 83-111.
[18] A. Stanbridge, D. Ewins, Modal testing using a scanning laser Doppler vibrometer, Mechanical systems and signal processing, 13 (1999) 255-270.

[19] P. Castellini, M. Martarelli, E.P. Tomasini, Laser Doppler Vibrometry: Development of advanced solutions answering to technology's needs, Mechanical Systems and Signal Processing, 20 (2006) 1265-1285.
[20] A. Cigada, P. Mazzoleni, E. Zappa, Vibration monitoring of multiple bridge points by means of a unique vision-based measuring system, Experimental Mechanics, 54 (2014) 255-271.
[21] P. Mazzoleni, E. Zappa, Vision-based estimation of vertical dynamic loading induced by jumping and bobbing crowds on civil structures, Mechanical Systems and Signal Processing, 33 (2012) 1-12.
[22] F. Cheli, P. Mazzoleni, M. Pezzola, E. Ruspini, E. Zappa, Vision-based measuring system for rider's pose estimation during motorcycle riding, Mechanical Systems and Signal Processing, 38 (2013) 399-410.
[23] D. Feng, T. Scarangello, M.Q. Feng, Q. Ye, Cable tension force estimate using novel noncontact vision-based sensor, Measurement, 99 (2017) 44-52.
[24] D. Feng, M.Q. Feng, E. Ozer, Y. Fukuda, A vision-based sensor for noncontact structural displacement measurement, Sensors, 15 (2015) 16557-16575.
[25] D. Feng, M.Q. Feng, Experimental validation of cost-effective vision-based structural health monitoring, Mechanical Systems and Signal Processing, 88 (2017) 199-211.
[26] J. Baqersad, P. Poozesh, C. Niezrecki, P. Avitabile, Photogrammetry and optical methods in structural dynamics–A review, Mechanical Systems and Signal Processing, (2016).
[27] X. Xie, D. Zeng, J. Li, J. Dahl, Q. Zhao, L. Yang, Tensile Test for Polymer Plastics with Extreme Large Elongation Using Quad-Camera Digital Image Correlation, in, SAE Technical Paper, 2016.
[28] J. Li, X. Xie, G. Yang, B. Zhang, T. Siebert, L. Yang, Whole-field thickness strain measurement using multiple camera digital image correlation system, Optics and Lasers in Engineering, 90 (2017) 19-25.
[29] P.L. Reu, T.J. Miller, The application of high-speed digital image correlation, The Journal of Strain Analysis for Engineering Design, 43 (2008) 673-688.
[30] P. Reu, Introduction to digital image correlation: Best practices and applications, Experimental Techniques, 36 (2012) 3-4.
[31] M.N. Helfrick, C. Niezrecki, P. Avitabile, T. Schmidt, 3D digital image correlation methods for full-field vibration measurement, Mechanical Systems and Signal Processing, 25 (2011) 917-927.
[32] C. Nonis, C. Niezrecki, T.-Y. Yu, S. Ahmed, C.-F. Su, T. Schmidt, Structural health monitoring of bridges using digital image correlation, Health Monitoring of Structural and Biological Systems, (2013) 869507.
[33] D.M. Revilock Jr, J.C. Thesken, T.E. Schmidt, Three-dimensional digital image correlation of a composite overwrapped pressure vessel during hydrostatic pressure tests, (2007).
[34] J. Baqersad, C. Niezrecki, P. Avitabile, Full-field dynamic strain prediction on a wind turbine using displacements of optical targets measured by stereophotogrammetry, Mechanical Systems and Signal Processing, 62 (2015) 284-295.
[35] J.G. Chen, N. Wadhwa, Y.-J. Cha, F. Durand, W.T. Freeman, O. Buyukozturk, Modal identification of simple structures with high-speed video using motion magnification, Journal of Sound and Vibration, 345 (2015) 58-71.
[36] J.G. Chen, A. Davis, N. Wadhwa, F. Durand, W.T. Freeman, O. Büyüköztürk, Video Camera–Based Vibration Measurement for Civil Infrastructure Applications, Journal of Infrastructure Systems, (2016) B4016013.

[37] J.G. Chen, N. Wadhwa, Y.-J. Cha, F. Durand, W.T. Freeman, O. Buyukozturk, Structural modal identification through high speed camera video: Motion magnification, in: Topics in Modal Analysis I, Volume 7, Springer International Publishing, 2014, pp. 191-197.
[38] Y. Yang, C. Dorn, T. Mancini, Z. Talken, G. Kenyon, C. Farrar, D. Mascareñas, Blind identification of full-field vibration modes from video measurements with phase-based video motion magnification, Mechanical Systems and Signal Processing, 85 (2017) 567-590.
[39] Y. Yang, C. Dorn, T. Mancini, Z. Talken, S. Nagarajaiah, G. Kenyon, C. Farrar, D. Mascareñas, Blind identification of full-field vibration modes of output-only structures from uniformly-sampled, possibly temporally-aliased (sub-Nyquist), video measurements, Journal of Sound and Vibration, 390 (2017) 232-256.
[40] P. Poozesh, A. Sarrafi, Z. Mao, P. Avitabile, C. Niezrecki, Feasibility of extracting operating shapes using phase-based motion magnification technique and stereo-photogrammetry, Journal of Sound and Vibration, (2017).
[41] A. Sarrafi, P. Poozesh, C. Niezrecki, Z. Mao, Mode extraction on wind turbine blades via phase-based video motion estimation, in: SPIE Smart Structures and Materials+ Nondestructive Evaluation and Health Monitoring, International Society for Optics and Photonics, 2017, pp. 101710E-101710E-101712.
[42] P. Poozesh, K. Aizawa, C. Niezrecki, J. Baqersad, M. Inalpolat, G. Heilmann, Structural health monitoring of wind turbine blades using acoustic microphone array, Structural Health Monitoring, (2016) 1475921716676871.
[43] S. Baker, I. Matthews, Lucas-kanade 20 years on: A unifying framework, International journal of computer vision, 56 (2004) 221-255.
[44] B.K. Horn, B.G. Schunck, Determining optical flow, Artificial intelligence, 17 (1981) 185-203.
[45] J. Javh, J. Slavič, M. Boltežar, The subpixel resolution of optical-flow-based modal analysis, Mechanical Systems and Signal Processing, 88 (2017) 89-99.
[46] A. Sarrafi, P. Poozesh, Z. Mao, A Comparison of Computer-Vision-Based Structural Dynamics Characterizations, in, Springer, International Modal Analysis Conference (IMAC) 2017, 2017.
[47] D.J. Fleet, A.D. Jepson, Computation of component image velocity from local phase information, International journal of computer vision, 5 (1990) 77-104.
[48] D. Fleet, Y. Weiss, Optical flow estimation, in: Handbook of mathematical models in computer vision, Springer, 2006, pp. 237-257.
[49] N. Wadhwa, Revealing and analyzing imperceptible deviations in images and videos, in, Massachusetts Institute of Technology, 2016.
[50] N. Wadhwa, M. Rubinstein, F. Durand, W.T. Freeman, Phase-based video motion processing, ACM Transactions on Graphics (TOG), 32 (2013) 80.
[51] I. Fogel, D. Sagi, Gabor filters as texture discriminator, Biological cybernetics, 61 (1989) 103-113.